\documentclass[10pt]{bmc_article}

\usepackage{cite} 
\usepackage{ifthen}  
\usepackage{multicol}   
\usepackage{bibunits}
\usepackage{times} 

\usepackage{mathrsfs}
\usepackage{amsmath}
\usepackage{amssymb}
\usepackage{amsthm}
\usepackage{color,graphicx}
\newcommand{\ud}{\,\mathrm{d}}
\definecolor{newtext}{rgb}{.5, .5, 0}
\usepackage{appendix}
\usepackage[T1]{fontenc}
\usepackage{multirow}
\usepackage{booktabs}
\usepackage{caption}
\usepackage[geometry]{ifsym}

\newcommand{\rev}[1]{\textcolor{black}{#1}}

\newboolean{publ}

\newenvironment{bmcformat}{\begin{raggedright}\baselineskip20pt\sloppy\setboolean{publ}{false}}{\end{raggedright}\baselineskip20pt\sloppy}

\begin{document}
\begin{bmcformat}

\defaultbibliography{DnaA_bibliography}

\defaultbibliographystyle{bmc_article}

\title{DnaA and the timing of chromosome replication in
\textit{Escherichia coli} as a function of growth rate}
 

\author{Matthew AA Grant$^1$%
\email{Matthew AA Grant - maag2@cam.ac.uk}%
      \and
      Chiara Saggioro$^2$%
\email{Chiara Saggioro - csaggior@lbpa.ens-cachan.fr }%
      \and
         Ulisse Ferrari$^3$%
         \email{Ulisse Ferrari - ulisse.ferrari@gmail.com}%
      \and
      	Bruno Bassetti$^{4,5}$%
      	\email{Bruno Bassetti - Bruno.Bassetti@mi.infn.it}%
      \and
      	Bianca Sclavi$^2$ %
      	\email{Bianca Sclavi - sclavi@lbpa.ens-cachan.fr}%
       and
     Marco Cosentino Lagomarsino\correspondingauthor $^{6,7,4}$%
      	\email{Marco Cosentino Lagomarsino\correspondingauthor  -
          marco.cosentino-lagomarsino@upmc.fr} %
      }


      \address{%
\iid(1) BSS Group, Department of Physics, University of Cambridge, JJ Thomson Avenue, Cambridge,        CB3 0HE, UK\\  
\iid(2) LBPA, UMR 8113 du CNRS, Ecole Normale Sup\'{e}rieurede Cachan,
61 Avenue du
Pr\'{e}sident Wilson, 94235 CACHAN, France \\
\iid(3) Dip. Fisica, Universit\`a ``Sapienza'', and IPCF-CNR, UOS Roma
Piazzale A. Moro
2, I-00185, Rome, Italy \\
\iid(4) Universit\`a degli Studi di Milano, Dip.  Fisica.  Via
Celoria 16, 20133 Milano, Italy \\
\iid(5) I.N.F.N. Milano, Italy\\
\iid(6) G\'enophysique / Genomic Physics Group, UMR7238 CNRS "Microorganism Genomics" \\
\iid(7) University Pierre et Marie Curie, 15 rue de l'\'Ecole de M\'edecine, 75006 Paris, France 
}%

\maketitle


\begin{abstract}
  \paragraph*{Background:}In \emph{Escherichia~coli}, overlapping
  rounds of DNA replication allow the bacteria to double in faster
  times than the time required to copy the genome. The precise timing
  of initiation of DNA replication is determined by a regulatory
  circuit that depends on the binding of a critical number of
  ATP-bound DnaA proteins at the origin of replication, resulting in
  the melting of the DNA and the assembly of the replication
  complex. The synthesis of DnaA in the cell is controlled by a
  growth-rate dependent, negatively autoregulated gene found near the
  origin of replication. Both the regulatory and initiation activity
  of DnaA depend on its nucleotide bound state and its availability.
  \paragraph*{Results:} In order to investigate the contributions of
  the different regulatory processes to the timing of initiation of
  DNA replication at varying growth rates, we formulate a minimal
  quantitative model of the initiator circuit that includes the key
  ingredients known to regulate the activity of the DnaA protein. This
  model describes the average-cell oscillations in DnaA-ATP/DNA during
  the cell cycle, for varying growth rates. We evaluate the conditions
  under which this ratio attains the same threshold value at the time
  of initiation, independently of the growth rate.  
\paragraph*{Conclusions:}
We find that a quantitative description of replication initiation by
DnaA must rely on the dependency of the basic parameters on growth
rate, in order to account for the timing of initiation of DNA
replication at different cell doubling times.
We isolate two main possible scenarios for this, depending on the
roles of DnaA autoregulation and DnaA ATP-hydrolysis regulatory
process.
One possibility is that the basal rate of regulatory inactivation by
ATP hydrolysis must vary with growth rate.
Alternatively, some parameters defining promoter activity need to be a
function of the growth rate. In either case, the basal rate of gene
expression needs to increase with the growth rate, in accordance with
the known characteristics of the \emph{dnaA} promoter.
Furthermore, both inactivation and autorepression reduce the amplitude
of the cell-cycle oscillations of DnaA-ATP/DNA. 
\end{abstract}

\ifthenelse{\boolean{publ}}{\begin{multicols}{2}}{}


\section*{Background}

The coordination of DNA replication with cell division in
\emph{E.~coli} is a classic problem of bacterial
physiology~\cite{Meselson1958}. It is connected with the control of
the bacterial DNA replication and cell division cycle as a function of
the growth rate, and it is an essential component for evolutionary
adaptation to fast-growing conditions~\cite{Rocha2004}. It is also a
classic problem for biological
modeling~\cite{Sompayrac1973,Margalit1984,Hansen1991,Browning2004,Nilsson2008}.
The main outstanding questions have to do with the characterization of
the network of regulatory interactions by which cells determine the
timing of initiation and limit it to once per cell cycle.

The theoretical foundations for understanding chromosome replication
initiation in \textit{E.~coli} were set by Cooper and Helmstetter
\cite{Cooper1968}, by showing that the time taken for a single
chromosome to be replicated ($C$ period) and the time period between
completion of chromosome replication and the following cell division
($D$ period) were approximately constant for a cell doubling time of
less than one hour~\cite{Michelsen2003}. The same work also introduced
the idea of overlapping rounds of chromosome replication, where a
round of replication can be initiated while an existing round of
replication is still proceeding (Figure~\ref{fig:fig1}). This
mechanism allows \textit{E.~coli} to grow with a doubling time faster
than the time required to copy its genome. The question then arises of
how the cell determines when to initiate DNA replication and how this
is coupled to the growth rate.
%

In 1968, Donachie calculated that the correct timing would be
guaranteed by a constant ratio of the cell size at the moment of
initiation (termed the `initiation mass') and the number of
\textit{oriC} in the cell~\cite{Donachie1968}.
Direct measurements of this ratio or of the initiation mass from cell
population are difficult. Thus, whether this ratio is effectively a
constant in cells with doubling times lower than one hour is to some
extent an open
question~\cite{Wold1994,Cooper1997,Boye2003,Bates2005a}.  It was then
proposed and debated that the amount of an initiating factor
accumulating with the cell's mass could reach a threshold value
resulting in activation of the origin(s).
The DnaA protein has been shown to possess the basic characteristics
necessary to act as such an
initiator~\cite{Loebner-Olesen1989,Hansen1991a}. Several monomers of
DnaA bind cooperatively to \textit{oriC} and induce DNA melting
required for assembly of the replication forks
\cite{Kornberg2005,Speck2001,Messer2002}. Its level of expression
increases with the growth rate \cite{Chiaramello1989} to result
approximately in a constant amount of total DnaA per cell
\cite{Hansen1991a}. DnaA overexpression results in earlier initiation,
and when it is depleted it results in delayed
initiation~\cite{Atlung1987,Bremer1985}.
%
%

In addition, DnaA exists under two forms, ATP or ADP bound. The first
is required for activation of the origin, thus it is usually called
the active form~\cite{Sekimizu1988}.
After the replication of \textit{oriC}, DnaA-ATP becomes converted to
DnaA-ADP in a process known as `RIDA' (\underline{R}egulatory
\underline{I}nactivation of \underline{D}na\underline{A}). RIDA is
mediated by the Hda protein and the beta clamp subunit of the
replisome and requires active replication forks
\cite{Donachie2003,Kurokawa1999}.
The hydrolysis of the ATP in a DNA-replication dependent manner
decreases the activity of the protein after initiation has taken
place, thus reducing the probability that a new initiation event will
occur within the same cell cycle~\cite{Donachie2003,Kurokawa1999}.  At
the same time the synthesis of new DNA creates new DnaA binding sites
that can titrate DnaA from the
origin~\cite{Ogawa2002,Morigen2001,Hansen1991}.

Other processes can contribute to prevent reinitiation within the same
cell cycle, such as the binding of the SeqA protein to the newly
replicated, hemimethylated DNA
\cite{Donachie2003,vonFreiesleben2000,Boye1996}. It is believed that
an ``eclipse'' period where reinitiation is not possible allows a
buffer time for the other processes such as RIDA to take effect and
thus for the levels of DnaA-ATP to decrease below the critical level
for initiation.  Several GATC sites are also found at the \emph{dnaA}
promoter but their effect on the timing of initiation remains to be
established \cite{Zaritsky2007,Kedar2000}.
Finally, a set of proteins have been shown to either inhibit or
enhance the activity of DnaA at the origin. These are for the most
part abundant nucleoid proteins such as FIS, HU and IHF, that may play
a regulatory role as a function of changes in the growth
phase~\cite{Polaczek1997}.

DnaA-ATP binding to the origin must determine the timing of initiation
for a range of growth rates and thus in the presence of increasing
genome amounts (providing non-specific binding sites).  Thus, the
amount of DnaA-ATP per cell needs to increase  with the
decrease in doubling time.  The \textit{dnaA} gene is found next to the origin
on the chromosome, resulting in the gene copy-number increasing with
the number of origins.  In addition, the expression of the \textit{dnaA} gene
is growth
rate-dependent~\cite{Chiaramello1989,Polaczek1990,Hansen1991a}. The
\textit{dnaA} promoter region contains multiple binding sites for DnaA with
differential affinity and specificity for the ATP- and ADP-bound forms
of the protein and has been shown to be autorepressed by DnaA-ATP but
not DnaA-ADP~\cite{Speck2001}. Consistent with this negative
autoregulation, the artificial addition of DnaA boxes in the cell
results in an increase in gene expression \cite{Hansen1987,
  Christensen1999,Morigen2001,Kitagawa1996} and inhibition of the RIDA
process or the presence of a mutant form of DnaA insensitive to RIDA
(DnaAcos) results in a decrease in the level of DnaA protein in the
cell \cite{Riber2006,Katayama1994}.
Finally, DnaA is a transcription factor for a set of genes involved in
regulation of DNA replication~\cite{Messer2002}
and it could thus act as a reporter of the DNA replication state of
the cell in order to maintain the correct stoichiometry of the DNA
replication regulatory factors at varying growth rates and in response
to perturbation to the movement of the replication forks~\cite{Olliver2010}.

It has previously been proposed that the presence of both
autoregulation and RIDA contributes to increased robustness of the
initiation regulatory network upon
perturbations~\cite{Riber2006,Donachie2003}. In this work, we aim to
determine the relative roles of of these two regulatory processes in
the control of the timing of initiation with changing growth rate.
We begin from the elements provided by the Cooper and Helmstetter
model in order to estimate the initiation time at different growth
rates.
The two main assumptions are that initiation of DNA replication is
determined by a critical amount of DnaA-ATP per non-specific site on
the genome and that this threshold value remains constant as a
function of growth rate.
On the other hand, the cellular and metabolic parameters can change
with growth rate and have an impact on the DnaA circuit.  In order to
understand this, we use information from systematic studies of
cellular changes with growth rate~\cite{Bremer1996,Klumpp2008}.
Finally, the volume of the cell is assumed to be a less relevant
background as a reservoir of DnaA-ATP than the number of non-specific
binding sites on the DNA~\cite{Bintu2005}.

The resulting equations describe, via a continuous change in parameter
values with growth rate, the oscillations in DnaA-ATP per non-specific
site and the attainment of a constant threshold as a function of
growth rate. 
This shows that the circuit performing the timing of replication
initiation must encode subtle information on the bacterial
physiological state through the growth rate dependence of the
parameters.
This analysis also allows us to define a few scenarios consistent with
the available experimental knowledge and to make testable predictions
on the relative roles of DnaA autorepression and of the RIDA process
at different growth rates.
We use this model to elucidate the reciprocal roles of the known
factors affecting DnaA activity in \emph{E.~coli}, namely that DnaA
expression is dependent on transcriptional autoregulation, and that
its ATP-ase activity is coupled with the activity of the advancing
replication forks (RIDA).
The results show that a working system can still be produced in the
absence of RIDA or DnaA autoregulation.
Moreover, both RIDA and autorepression contribute to a decrease in the
amplitude of the cell-cycle oscillations in DnaA-ATP. RIDA has a
larger effect at the faster growth rates while negative regulation has
a larger effect at slower growth.

  
\section*{Methods}

\textbf{Assumptions of the model.}
The model consists of a set of Ordinary Differential Equations (ODEs)
describing DnaA-ATP production by the expression of the \textit{dnaA} gene.
It is built on two basic assumptions. The first is based on the
evidence that a specific number of DnaA-ATP molecules need to bind on
oriC in order to create a replication bubble. Following the standard
thermodynamic model of protein-DNA binding
\cite{Bintu2005,Shea1985,Ackers1982}, the probability of this event is
dependent on the number of DnaA-ATP molecules that are bound to the non-specific
sites along the chromosome. 
\rev{These are low-affinity sites compared to titration sites, but
  the affinity is high enough so that the protein spends most of its
  time bound to the genome.  These sequence-independent interactions
  are typical of DNA-binding proteins. As a consequence, the
  simplifying assumption is usually made~\cite{Bintu2005} that the key
  molecular players (RNAP and TFs) are bound to the DNA either
  specifically or non-specifically.  Simply stated, this is just an
  implementation of the known fact that DNA-binding proteins, besides
  binding tightly to their target sequences, are generally ``sticky''
  for DNA, in a sequence-independent manner.  }
This implies that the timing of DNA replication initiation in the cell
is determined by the ratio DnaA-ATP to non-specific binding sites on
DNA (which in turn must be proportional to the total DNA length of the
chromosome(s) in the cell and effectively results in the computation
of the amount of DNA per cell). Note that in this case the volume of
the cell is not taken into consideration since it is not the change in
concentration of DnaA or of DNA that determines the initiation
time. The same will apply to the binding of DnaA and RNA polymerase to
the \textit{dnaA} promoter (see below).
The second assumption is that at the time of initiation this ratio
will be the same, independently of the growth rate and thus the number
of origins.

We ask how the parameters of this model must vary in order for this
assumption to hold in the range of doubling times between 20 and 60
minutes.
In the absence of autoregulation, the only factor that contributes to
a decrease in the ratio of DnaA-ATP to non-specific binding sites is
the increase in DNA after DNA replication has begun.
The complete model also includes the autoregulation of DnaA expression
by DnaA binding to its own promoter~\cite{Speck2001} and DnaA-ATP
transformation into DnaA-ADP through the RIDA
process~\cite{Riber2006,Camara2005}. It is assumed that this ODE
description is applicable to a single average-cell on time scales
shorter than the length of the cell cycle. This hypothesis could be
challenged for the shortest observable doubling times, but the
formulation of the model is dictated by maximizing simplicity.
The values of the parameters (attributed to a specific value of the
growth rate) are all taken or estimated from the available
experimental measurements. They are shown in
Table~\ref{tab:initial_values}, together with the sources.

\subsection*{Formulation of the model}
    	
\paragraph*{Timing of replication.}
We take into consideration the situation where the cell cycle repeats
itself identically i.e. balanced, exponential growth. Following Cooper
and Helmstetter, at a time $C+D$ after the initiation time, the cell
divides, i.e. that time must be an integer multiple of the doubling
time. Thus, if $\tau $ is the doubling time of the cell and $X$ is the
initiation time, then we must have
\begin{equation}
  X+C+D=(n+1)\tau \ ,
  \label{eq:CH}
\end{equation}
where $n$ is the integer number of times that $\tau $ divides
$C+D$. $n$ can be viewed as the number of overlapping rounds of
replication, and $2^n$ is the number of origins. Thus, this
equation reflects the phenomenon of overlapping replication
rounds. Figure~\ref{fig:fig1} shows how $X$ varies with the
doubling time ($\tau $) of the cell. Defining $Y$ as the time at which
the chromosome completes replication, we have
	
\begin{equation}
  Y=\tau-D \ .
  \label{eq:Y}
\end{equation}

\paragraph*{Promoter term.}
The activity of the \textit{dnaA} promoter is the source term for DnaA-ATP. We
describe it by the standard thermodynamic model first used by Shea and
Ackers \cite{Bintu2005, Shea1985, Ackers1982}. We denote the promoter
term (the number of DnaA-ATP synthesised per unit time) as $Q$, the
number of non-specific binding sites on the chromosome as $N_{NS}$ and
the number of RNA polymerase (RNAP) molecules as $P$. Furthermore, we
denote the number of DnaA-ATP molecules as $A_-$. We then use the
assumption that the number of non-specific binding sites is
proportional to the length of DNA in the cell (which we write as
$\Lambda$), i.e. $ \Lambda=\kappa N_{NS}$.
 
Thus the expression for the rate of transcription at the \textit{dnaA} promoter
can be written as (see Additional File 1,~\ref{sec:promoter})
\begin{equation}
Q	= \frac{k_A
  \Theta}{1+c_1\frac{\Lambda}{P}+c_2\frac{A_-}{P}}  \label{eq:Q}  
\end{equation}
where $k_A$ is the basal rate of transcription of the \textit{dnaA} promoter,
$\Theta(t,\tau)$ is the number of \textit{dnaA} promoters (and genes) in a
given cell at a given time, and the remaining factor is the
probability of RNAP binding to the promoter. The parameters $c_1$ and
$c_2$ depend on the binding energies $\Delta \epsilon_{pd}$ and
$\Delta \epsilon_{ad}$ of RNAP and $A_-$ respectively to their
promoter binding sites. The binding energies are determined from the
ratio of specific vs non-specific binding affinities. 

\begin{align}
c_1	&=\frac{e^{\Delta \epsilon_{pd}/k_BT}}{\kappa} \nonumber \\
c_2	&=e^{(\Delta \epsilon_{pd}-\Delta \epsilon_{ad})/k_BT}.  
  \label{eq:binders}
\end{align}
where the exponential terms are Boltzmann weights.  $c_2=0$ if the
promoter is not autorepressed. A version of the promoter where DnaA
binding to its sites is cooperative is described in Additional File 1.

\paragraph*{RIDA term.}
This term reflects the number of DnaA-ATP molecules that are converted
to DnaA-ADP molecules per unit time by the RIDA process. As discussed
in the introduction, RIDA is a process that takes place at the
replication forks during DNA synthesis. We assume that the rate of
conversion $k_R$ takes the same value at each replication fork. The
number of pairs of replication forks at a given time, $\mathcal{F}(t)$, depends on which of $X$ and $Y$ is larger.
		
For $X<Y$:
\begin{equation}
  \mathcal{F}(t)= \left\{
    \begin{array}{rl}
      2^n-1 & \text{if } 0<t<X \\
      2\cdot2^n-1& \text{if } X<t<Y \\
      2(2^n-1)& \text{if } Y<t<\tau 
    \end{array} \right.
\end{equation}

and for $X>Y$:
\begin{equation}
  \mathcal{F}(t)= \left\{
    \begin{array}{rl}
      2^n-1 & \text{if } 0<t<Y \\
      2^n-2& \text{if } Y<t<X \\
      2(2^n-1)& \text{if } X<t<\tau 
    \end{array} \right.
\end{equation}

\rev{(note that this equation is intrinsically discrete since it
  relates to the physical number of replication forks)} and so the
conversion from DnaA-ATP to DnaA-ADP takes place at a rate

\begin{equation}
  k_R \mathcal{F}(t).
  \label{eq:RIDA}
\end{equation}

This leads to the following differential equations for DnaA-ATP
(denoted $A_-$) and DnaA-ADP (denoted $A_+$)

\begin{align}
  \frac{\partial A_-}{\partial t}	&=\frac{k_A
\Theta}{1+c_1\frac{\Lambda}{P}+c_2\frac{A_-}{P}}-k_R \mathcal{F}\label{eq:A_minus_eq} \\
  \frac{\partial A_+}{\partial t} &=k_R \mathcal{F}.
  \label{eq:A_plus_eq}
\end{align}

\paragraph*{Term for the growth of the chromosome.}
The growth of the chromosome is controlled by the replication
forks. Defining the rate of DNA synthesis of each pair of replication
forks
as $k_\Lambda$, we can write
\begin{equation}
  \frac{\partial \Lambda}{\partial t}=k_\Lambda \mathcal{F}.
  \label{eq:lengthgrowth}
\end{equation}
Assuming that $k_\Lambda$ is constant, and normalizing so that
$\Lambda=1$ is the length of one full chromosome, we have $k_\Lambda=
1/C$.

\paragraph*{Main equation.}
Figure \ref{fig:fig2} summarizes the ingredients of the model.
Defining
$r=\frac{A_-}{\Lambda}$ and combining \eqref{eq:A_minus_eq} and
\eqref{eq:lengthgrowth} we obtain the equation

\begin{equation}
\frac{\partial r}{\partial
t}=\frac{1}{\Lambda}\left(\frac{\Theta
k_A}{1+c_1\frac{\Lambda}{P}+c_2\frac{\Lambda}{P}r}-(k_R+rk_\Lambda)\mathcal{F}\right).
  \label{eq:full}
\end{equation}

This equation describes the dynamics of the variable
$r=\frac{A_-}{\Lambda}$ which we suggest is a suitable candidate for
the initiation potential since a specific number of DnaA-ATP molecules
is needed to be available to bind to the origin in order to induce DNA
melting, as described above.
Note that usually such dynamic equations are written in terms of
volume, thinking of averages over cell populations on time-scales
longer than a cell cycle. We assume that the (time-varying) background
of genomic binding sites is the relevant variable in a Shea-Ackers
type model, extending to a single cell cycle the approach normally
used for longer time scales~\cite{Bintu2005,Shea1985,Ackers1982}. We
also assume that the volume can be treated as a weak perturbation,
which we neglect here.  In other words, the various molecules of
interest (RNA Polymerase, DnaA) are partitioned between the specific
and non-specific binding sites on the chromosome. Furthermore, the
most important factor that determines the probability of binding to a
given promoter is the absolute number of protein molecules relative to
the absolute number of these binding sites, rather than the amount of
protein per cell volume, which in comparison does not change
significantly, and it can thus be neglected to a good
approximation. Thus, this assumption means that one need not track the
volume of the cell, only the number of non-specific binding sites in
the cell at a given time. This idea is discussed further in
Additional File 1,~\ref{sec:promoter}.
Here we assume that the initiation potential, $r$, always reaches the
same value at $t=X$ independently of the growth rate and we ask how
the parameters of this model must vary in order for this assumption to
hold in the range of doubling times between 20 and 60 minutes. In the
following, we will first establish that such a constant threshold
cannot be obtained by a model with fixed parameters and then study the
possible scenarios where different subsets of parameters are allowed to
vary.



\paragraph*{Main Assumptions. }
The model relies on the following further
assumptions~\cite{Donachie2003}: (i) All newly synthesised DnaA is
immediately bound to ATP, due to the relative abundance of ATP in the
cell compared to ADP and the high affinity of DnaA for ATP. (ii)
DnaA-ADP is only created by conversion from DnaA-ATP by the RIDA
process, when it is present. (iii) The probability of DnaA being bound
to its sites at the origin or on the promoter, in the case of the
presence of autorepression, is given by its thermodynamic equilibrium
value. The same assumption holds for the binding of RNA polymerase to
a particular promoter. This means that we assume that the rate for
transcription initiation is much slower than the rates for RNAP
binding and unbinding from the promoter. (iv) The rate of
\textit{dnaA} gene expression is proportional to the equilibrium
probability that RNAP is bound to the \textit{dnaA} promoter.  (v) We
do not consider translation directly and thus there is no time delay
from transcription to protein production since the addition of this
feature did not affect the result of the model (see Results). (vi) The
number of non-specific binding sites on the DNA in each cell for both
RNAP and DnaA is proportional to the total length of DNA in the
cell~\cite{Hippel1986}. (vii) The number of RNAP molecules in the
cell, $P$, grows exponentially from cell birth to cell division,
corresponding to the hypothesis of constant concentration and
exponentially growing cell size~\cite{Godin2010}. A linear growth can
also be used, however the dynamics of the model do not differ
significantly between these two cases.

\paragraph*{Numerical integration. }
The non-linearity of the main equation~\eqref{eq:full} necessitates
the use of a numerical method of integration. We used a custom C++
implementation of the fourth-order Runge-Kutta method. The equation
was integrated for values of the cell doubling time, $\tau$, in the
range $21 \text{mins}\le\tau\le60 \text{mins}$.

In order to test for the constant threshold condition, a
transformation was performed by integrating the equation for $\tau=21$
mins and using the value of $r$ at initiation $t=X$ as the imposed
threshold for the other values of the doubling time. Thus it was
important to estimate, to a good degree of accuracy, values for the
parameters at a doubling time of 21 mins. These values appear in
Table~\ref{tab:initial_values}. The parameter values are either taken
to be constant (independent of cell doubling time) or are allowed to
change and obtained as a consequence of the transformation (see
Table~\ref{tab:scenarios2} for whether a parameter is constant or
allowed to vary in a given situation).

\section*{Results}

\subsection*{A fixed set of parameters gives a varying initiation
  threshold with increasing growth rate.}

We first describe the behaviour of the model with a fixed parameter
set.
The ratio $r=\frac{A_-}{\Lambda}$ increases from the time of birth of
the new cell. This can be interpreted as the accumulation of the
`initiation potential'. At initiation ($t=X$), $r$ peaks (at the
`initiation potential' threshold) and then falls again due to the
increase in the number of non-specific binding sites.  When including
the RIDA process, the total RIDA rate also increases following
initiation, due to the higher number of active replication forks,
contributing to the decrease in the initiation potential.  However,
the value of $r$ at $t=X$ varies for the different values of the
doubling time $\tau$ (Figure~\ref{fig:fig3}A).  Thus, it appears that
this model, with fixed parameters, cannot give a constant threshold
that is reached at initiation in the range of growth rates considered
here.

This fact naturally leads to us consider a model in which the parameters
are able to vary with growth rate. Biologically, this is a natural
requirement, as one may well expect from previous observations of the
change in cellular components as a function of growth rate
\cite{Bremer1996,Klumpp2008,Scott2011,Scott2010}.

\subsection*{A constant threshold condition implies alternative
  scenarios of growth-rate dependency in the circuit architecture}

The condition of a constant DnaA-ATP/DNA threshold at the time of
initiation can be imposed by performing a mathematical transformation
on the model and verifying the implications of this for the values of
the parameters. The mathematical details of this transformation can be
found in Additional File 1,~\ref{sec:math_trans}.
The transformation yields a fixed threshold in $r$ that is reached at
initiation. This can be seen upon comparing the plots in
Figure~\ref{fig:fig3}A and B. In the latter, the value of $r(X)$ is
now the same at initiation for every cell doubling time. We estimate
all initial values of the parameters from the literature (Table 1) and
then we allow some of the parameters to change during the
transformation. Thus, the transformation procedure imposes a decision
on which parameters to fix and which to allow to change. This
determines different scenarios, as illustrated in
Figures~\ref{fig:fig4} and~\ref{fig:fig5} and also summarised in
Table~\ref{tab:scenarios2} (see also Additional File 1,~\ref{sec:math_trans} and
Additional File 1, Figure~\ref{fig:SF0}). \rev{These plots explicitly show
  the required parameter changes at varying growth rates.  }

\begin{enumerate}
\item In the first scenario, the RIDA rate (per replication fork),
  $k_R$, is chosen to remain independent of growth rate, and the other
  parameters are allowed to change.  The transformation fixes the
  scaling of $k_A$ with growth rate, and provides an equation that
  $c_1$ and $P$ must satisfy. Since the constraints reduce to one
  equation with two unknowns, there is a set of possible solutions for
  $c_1$ and $P$, and the scenario is underconstrained
  (see Additional File 1,~\ref{sec:math_trans}).
  One possibility is to fix one of the parameters to a particular
  trend, in turn determining the second parameter, determining
  different subscenarios (Additional File 1, Figure~\ref{fig:SF0}).
  \begin{enumerate}
  \item In the first of these sub-scenarios, 1a, the variation in the
    number of non-specifically bound RNAP in the cell, $P$, with growth
    rate is fixed \emph{a priori} by imposing the trend of a previous
    study~\cite{Klumpp2008}, which partitions RNAP into different
    classes, the RNAP bound on promoters and non-specifically bound,
    and determines the dependence of this partition with growth rate,
    (see Figure~\ref{fig:fig4}A). Imposing this trend determines the
    values for the binding affinity of RNAP to the promoter, which
    must vary with growth rate (Additional File 1,~\ref{sec:math_trans}).
  \item In sub-scenario 1b, the binding affinity of RNAP to the
    \emph{dnaA} promoter is chosen to remain independent of growth
    rate. Fixing this parameter constrains the remaining equation,
    thereby imposing a particular dependence on growth rate of the
    number of non-specifically bound RNAP, $P$, which turns out to be
    compatible with ref.~\cite{Klumpp2008} (see Additional File 1,~\ref{sec:math_trans}
    and Figure~\ref{fig:fig4}A).
  \end{enumerate}
\item In scenario 2, the binding affinity of both RNAP to the \emph{dnaA}
  promoter and of DnaA-ATP to its repression sites are chosen to
  remain independent of growth rate (in simulations, at the values
  shown in Table~\ref{tab:initial_values}), while the basal rate of
  transcription from the \textit{dnaA} promoter, $k_A$, the levels of free
  RNAP, $P$, and the RIDA rate (per replication fork), $k_R$ are all
  permitted to vary from their original values (see
  Figures~\ref{fig:fig4}A and~\ref{fig:fig5}B). This scenario is also
  feasible in absence of autorepression.
\end{enumerate}

In brief, two possible scenarios can result in a constant initiation
threshold for the model. In the first, the binding affinity of
DnaA-ATP to its repressor sites decreases with increasing cell
doubling time and in the second the RIDA rate increases with cell
doubling time. In both scenarios $k_A$ (the basal transcription rate)
must decrease with increasing cell doubling time.
It is important to note that $k_A$, the basal \emph{dnaA}
transcription rate, needs to vary in all scenarios with growth
rate. This term (and hence its variation) is independent from RNAP
availability (our $P$ term) and binding (our $c_1$ and $c_2$ terms).
It describes how quickly RNAP moves through a gene when
transcribing. The variation of this characteristic time with growth
rate can be associated with variations in DNA supercoling (see
Discussion).

In absence of RIDA ($k_R = 0$) or autorepression ($c_2 = 0$), the
transformation can still be performed. However, it implies that the
ratio $c_1 / P $ should remain constant with growth rate. Since $P$
varies~\cite{Klumpp2008}, this means that $c_1$ (which, for example,
could also vary through changes in supercoiling) would have to
compensate exactly for the changes in $P$ in order to keep a constant
threshold.
We have considered these further scenarios to be less probable,
because they may result in a less robust control due to an unlikely
fine-tuning of two parameters.

\subsection*{The resulting scenarios are compatible with available
  knowledge on RNAP availability and total DnaA expression}

Given these scenarios, we have asked whether the predicted parameter
variation with growth rate and the observables quantities produced by
the model were compatible with the measurements and observations
available in the literature.

Starting from the dependency of available RNAP with growth rate, this
is predicted and matched with available experimental data in
ref.~\cite{Klumpp2008}.  Scenario 1a assumes this dependency, and
therefore automatically accounts for this observation.
Figure~\ref{fig:fig4}A demonstrates how also scenarios 1b and 2 are
broadly compatible with the results of Klumpp and
Hwa~\cite{Klumpp2008} since the average levels of non-specifically
bound RNA polymerase decrease with increasing doubling time in all
cases.

We now turn to the changes in measured expression of DnaA (averaged
over a population) with growth rate. This can be measured by a
reporter gene technique.  Figure~\ref{fig:fig4}B shows how the model
predicts that the average expression of the $dnaA$ gene should change
with growth rate in all scenarios.  This appears to be independent of
the scenario chosen and is compatible with previous
findings~\cite{Chiaramello1989}. We have also performed our own
measurements, using a GFP reporter of the \textit{dnaA} promoter
encoded on a plasmid (and normalizing the result for plasmid and gene
copy number (Chiara Saggioro, Anne Olliver, Bianca Sclavi: Multiple levels of regulation in the growth rate dependence of DnaA expression, submitted), see also \rev{ Additional File 1,~\ref{sec:experimental_methods}}), confirming this agreement
(Figure~\ref{fig:fig4}B).

Finally, at fixed growth rate, in order to determine whether this
model reflects the main features of the regulatory network in the
cell, we reproduced some of the experimental perturbations described
in the literature. One of these experiments changed the rate of RIDA
by changing the level of expression of the gene encoding for the Hda
protein~\cite{Riber2006}, while others controlled the expression of
DnaA independently of the \textit{dnaA} promoter.
We found that a constant threshold can be obtained upon a 10 fold
change in the RIDA rate (Additional File 1, Figure~\ref{fig:SF8}).  In
order to determine how RIDA rate and autorepression strength influence
the activity of DnaA-ATP within a cell cycle in the model, we have
monitored the amplitude of the oscillations of the ratio
DnaA-ATP:total genome length.  The results show that increased
autorepression contributes to a smaller amplitude of DnaA-ATP within
the cell cycle (Additional File 1, Figure~\ref{fig:SF5}).  A decrease in
RIDA has the opposite effect, and is compensated for by an increase in
autorepression.
In either case the inclusion of these additional control factors
appears to result in smaller amplitude in the oscillation of DnaA-ATP
and a smaller variation in the average amount of DnaA-ATP per cell as
the growth rate is varied. This may be advantageous for the use of
DnaA as a transcription factor whose activity is responsive to changes in
the replication status of the cell, via the RIDA and titration
effects.
%
%

Finally, once we obtained a set of parameters that satisfies the
constant threshold constraint, we modified the RIDA rate while leaving
the other parameters unchanged (Additional File 1, Figure~\ref{fig:SF9}), attempting to
reproduce the effect of under or over expressing the Hda protein, as
in the experiments by Riber \emph{et al.}~\cite{Riber2006}.  In our
model, this results in a change of the threshold value as a function
of growth rate, and more significantly at slow growth. At faster RIDA
rate, the threshold value is higher at slow growth, while the opposite
is observed when RIDA rate is decreased.

These compatibility tests give positive results, but do not allow us to
distinguish between the two scenarios.  We have explored the
literature for tests of dependency of the RIDA rate with growth rate,
and have found no evidence of this, which lead us to consider scenario
1, where the RIDA rate is constant as the main one.


\subsection*{Model Variants}

In order to gain confidence that the conclusion (that
the parameters need to vary with growth rate) is not a consequence of
the restricted set of biological ingredients included in this model,
we considered some additional model variants, including some of the
known factors that may influence the timing of replication initiation.

\begin{enumerate}


\item \textbf{Delay in the synthesis of DnaA-ATP}. We introduced a
  delay, representing the time necessary to obtain an active DnaA
  molecule from the binding of the RNA polymerase to the \textit{dnaA} promoter
  to the end of translation.  This delay, however, does not produce a
  significant effect in imposing a DnaA-ATP threshold at initiation,
  suggesting that translation delay might not have a predominant role
  in controlling the timing of initiation (see Additional File 1, Figure~\ref{fig:SF4}B).

\item \textbf{Cooperativity of autorepression}.  Cooperativity of
  autorepression affects the growth rate dependence of gene
  expression~\cite{Klumpp2008}.
  Additional File 1, Figure~\ref{fig:SF4}C shows by simulation that
  the presence of cooperativity alone cannot explain a constant
  threshold. Thus it is still necessary to impose a constant threshold
  on the model using a transformation such as that described in
  Additional File 1,~\ref{sec:math_trans}. However, in general one cannot make a
  translation and scaling and keep the promoter in the same form (see
  also Additional File 1,~\ref{sec:complex_promoter}), and the transformation itself
  poses complex constraints on the possible regulations. Thus, we
  decided to concentrate on the simpler non-cooperative
  model. Biologically, of the two high-affinity DnaA sites found at
  the \textit{dnaA} promoter, one matches exactly the consensus
  sequence, and has a higher affinity than the other~\cite{Speck2001},
  suggesting that at lower DnaA concentrations only one monomer could
  be bound.  We have verified that the transformation is possible with
  a promoter that would fit this profile, i.e. of the form
 \begin{displaymath}
 Q = \frac{k_A \Theta}
  { 1+c_1\frac{\Lambda}{P}
  \left(\frac{ 1 + \frac{r}{k_1} + \frac{r^2 \omega}{k_1
        k_2}}{1+\frac{r}{k_1}}  \right)  
  }  \ ,
\end{displaymath}
where the parameter $\omega$ represents the cooperativity, and where
$k_1$ and $k_2$ are the binding affinities of the two DnaA binding
sites, multiplied by the proportionality constant between $\Lambda$
and $N_{NS}$.
This description introduces two new parameters - the cooperativity and
the binding affinity of the second binding site.  Since these
parameters appear in the equations describing the transformation that
keeps the constant threshold, the scenarios become more
underconstrained in a non-essential way. Thus, while a more realistic
promoter model could be useful for future descriptions, we decided not
to pursue it here.

\item \textbf{The \textit{datA} locus \rev{and specific DnaA binding
      sites}}. We considered the effect that the presence of the
  \textit{datA} locus has on the model by introducing a site on the
  chromosome that binds up to 300 DnaA-ATP molecules immediately after
  initiation has taken place (since \textit{datA} is close to the
  origin on the chromosome, it is copied soon after initiation). We
  included one \textit{datA} site for each origin in the cell. The
  results indicate that the \textit{datA} locus might indeed prevent
  further initiations in a given cell cycle, since it titrates large
  numbers of DnaA-ATP molecules, effectively preventing them from
  binding at the origin (see Additional File 1, Figure~\ref{fig:SF4}A). However, this
  variant of the model fails to achieve a constant DnaA-ATP threshold
  at initiation at different growth rates if all parameters are kept
  constant. One can speculate that the large increase in \emph{datA}
  sites at faster growth rates would require a proportional increase
  in the rate of DnaA synthesis that cannot be solely provided by the
  increase in gene copy number.  It must be also noted that the
  \emph{datA} locus appears to be unnecessary to prevent reinitiation
  and limit initiation of to once per cell cycle~\cite{Morigen2005}.
  \rev{Additionally, we considered the effect of the reported $\approx
    300$ binding sites distributed around the
    chromosome~\cite{Roth1998}. These high-affinity sites would
    sequester DnaA in a similar way as the \emph{datA} locus, but
    proportionally to the genome amount, and thus effectively decrease
    its concentration. This is again insufficient to guarantee a
    constant initiation threshold and is equivalent to rescaling the
    RIDA rate (see Additional File 1,~\ref{sec:specific_binding}).  }

\item \textbf{The eclipse}.  No constraint for the eclipse period \rev{\cite{Donachie2003, vonFreiesleben2000, Boye1996, Zaritsky2007,Kedar2000}} was
  included in this model. This proved not to be necessary, probably
  because there is no delay between the attainment of the threshold
  and the initiation of DNA replication, thus avoiding the possibility
  that there is an `overshoot' of initiation potential before both
  RIDA and the increase of non-specific DNA can begin. On the other
  hand, the gene copy number immediately doubles upon initiation of
  DNA replication which results in a sudden increase in the number of
  DnaA-ATP synthesized per unit time. However, this does not result in
  an increase in the initiation potential due to the corresponding
  increase in the number of replication forks \rev{and thus on the RIDA rate and the number of non-specific titrating sites.}

\rev{
\item \textbf{DnaA-ATP Recycling regions}. Genomic recycling regions
  catalyzing the reconversion of DnaA-ADP into
  DnaA-ATP~\cite{Fujimitsu2009} have been recently discovered. The
  quantitative contribution of this process is not clear, but within
  our framework it makes sense to ask how this would affect the
  initiation threshold, as this process is, roughly, a correction to
  RIDA.  Precisely, assuming DnaA-ADP is not limiting, this would
  change the model by the substitution of $k_R$ with $k_R - \rho
  \Lambda/\mathcal{F} $ where $\rho$ is a fixed recycling rate. We
  verified that in the model this recycling term cannot by itself
  impose a constant threshold, while it can contribute to correcting
  quantitatively the effective RIDA rate (Additional File 1,~\ref{sec:DARS_considerations}).  }

\end{enumerate}

\rev{Since none of these model variants qualitatively changes the
  behaviour of the model with respect to attaining a constant
  initiation threshold, they were not included in the minimal model
  formulation, in order to avoid confusion and proliferation of
  parameters.  However, as shown by the variants explored above, the
  qualitative behaviour that the parameters of the model must vary
  with growth rate does not hold strictly for the minimal model only,
  but might be more general.}

\section*{Discussion}		

Standard models of bacterial regulatory circuits were adapted to
situations where the growth rate is fixed~\cite{Bintu2005,Alon2006}.
The notion that these quantitative descriptions must account for
bacterial physiology through the growth-rate dependent basic
partitioning of the cell physico-chemical components is now entering
the field of systems biology through a combination of new
work~\cite{Klumpp2008,Scott2010,Scott2011,Zaslaver2009} and
reconsideration of the
classics~\cite{Bremer1996,Cooper1968,Schaechter1958}.

The dependency of the basic parameters on growth rate can produce
notable effects on a genetic circuit, and complicates the standard
descriptions~\cite{Klumpp2009}.  In our case, the task is more
difficult, as the circuit under examination is active in \emph{determining}
some features of the bacterial physiology and not only affected by
them. Furthermore, on the technical level, one must produce a
time-dependent description the expression of DnaA over cell-cycles of
a range of durations.
Perhaps also for this reason, despite the fact that the regulation of
DNA replication has been a subject of intense study for over 50
years~\cite{Donachie2003,Katayama2010}, many questions remain open.
Given these obstacles, we have shown that, under a series of
simplifying hypotheses, a consistent mean-field description for the
DnaA / replication initation circuit is possible with varying growth
rate.

Our description includes the processes that are believed to be most
important for initiation of replication~\cite{Donachie2003}.  
In these respects, it is broadly compatible with previous modelling
approaches~\cite{Margalit1984,Hansen1991,Browning2004,Nilsson2008}. Its
originality lies in the minimality and in the attention given to
growth-rate dependency.
We focused on the minimal ingredients necessary in order for the basic
tenet that the ratio DnaA-ATP/DNA attains a constant threshold at
initiation to hold~\cite{Wang2011,Charbon2011}.  The validity of this
tenet is confirmed by the recent observations that initiation time is
not affected by adding an extra origin on the
chromosome~\cite{Wang2011} and on the compensatory mutations emerging
in Hda mutants~\cite{Charbon2011}.

We have defined the DNA replication initiation potential, determining
the (synchronous) timing of DNA replication, as the DnaA-ATP to DNA
ratio, $r$. Molecular titration has been shown to result in
ultrasensitive ``all or none'' responses~\cite{Buchler2008}, which
further justifies using $r$ as the threshold and could explain the
synchrony of initiation in cells containing \textit{oriC}
minichromosomes~\cite{Loebner-Olesen1999}.
We assume that its value at the time of initiation, $r(X)$, is
independent of the specific growth rate. 
The amount of DnaA-ATP at the time of initiation thus needs to
increase as a function of growth rate in order for $r(X)$ to remain
constant as a function of doubling time, and we found that
consequently, some of the model's parameter values must be allowed to
vary.
This assumption has not been verified directly. On the other hand, we
feel that our point of view would still be useful in case of a
growth-rate dependent $r(X)$, as it is unlikely that this dependence
would automatically match the dependence of all the other parameters.

We have defined two main scenarios in which different subsets of the
parameters are allowed to change.
In Scenario 1 the RIDA rate (per replication fork), $k_R$, is held
constant as a function of growth rate, but the binding affinities of
RNAP and DnaA-ATP to the DNA need to vary with growth rate
(note that in addition, there are two technical sub-scenarios to
  Scenario 1 due to the possibility of either fixing the growth rate
  dependence of $P$, the number of available RNA polymerase molecules
  a priori to the trend of ref.~\cite{Klumpp2008}(Scenario 1a) or
  allowing it to be free (Scenario 1b)).
In Scenario 2 the binding constants (c.f. $c_1$ and $c_2$) are
independent of growth rate but the RIDA rate, $k_R$, must vary. 
We have verified that both scenarios are consistent with the
eperimentally tested predictions of RNAP availability with growth
rate~\cite{Klumpp2008} and with previous
measurements~\cite{Chiaramello1989} and our own experimental
evaluation of total DnaA expression (Chiara Saggioro, Anne Olliver, Bianca Sclavi: Multiple levels of regulation in the growth rate dependence of DnaA expression, submitted), and also with
a number of ``in silico mutations'' inspired by the available
literature~\cite{Riber2006,Donachie2003}.  Thus, the scenarios appear
as possibilities that are testable, but for the moment remain open.
\rev{Note that the property that the initiation threshold holds
  constant with respect to growth rate changes \emph{is not} related
  to the specific set of parameters we used, or any set of parameters.
  Our analysis shows that in general, for any fixed parameter set at a
  given growth rate, a transformation is necessary in order to keep
  the threshold constant while moving to another growth rate. In order
  to provide specific examples, we have produced plots in the style of
  those in Figure~\ref{fig:fig5}, with different curves corresponding
  to choosing different values of the initial input parameters. These
  demonstrate that the qualitative behaviour of the transformation is
  independent of these parameters (Additional File 1, Figures
  \ref{fig:robustness1}, \ref{fig:robustness2} and
  \ref{fig:robustness3}). This exercise is also important to show that
  the parameter changes with growth rate are not numerically
  negligible for empirically plausible parameters, so that the
  question of keeping the initiation threshold constant is not purely
  academic.
}

It is then interesting to ask which of these scenarios is more
reasonable considering the known biological processes.
We speculate that scenario 2 is less likely, since until now there is
no evidence pointing to a possible change in the intrinsic RIDA rate
as a function of growth rate. The DnaA-related protein Hda
(\underline{H}omologous to \underline{D}na\underline{A}) mediates this
process~\cite{Katayama2010}. Experiments with mutants over- and
underexpressing Hda~\cite{Riber2006}, with corresponding increases and
decreases in the RIDA rate, suggest a possible mechanism by which the
$k_R$ term in the equations could vary by a growth rate-dependent
expression of the Hda gene. There may also be other, as yet unknown,
factors that affect the growth rate dependence of the RIDA rate.
Alternatively, we can speculate that the decrease in the rate of RIDA
with growth rate could be caused effectively by the action of the
reverse process of DnaA-ATP recycling by the recently discovered
recycling regions~\cite{Fujimitsu2009}. \rev{ Figure~\ref{fig:fig5}b
  shows that the RIDA rate should increase with cell cycle time and
  thus decrease with growth rate.  This growth rate increase causes
  overlapping replication rounds, and thus higher chromosome copy
  number. Since more recycling regions are present there is more
  recycling, \emph{i.e.} a decrease in the effective RIDA rate,
  compatibility with the requirement imposed by our results.
  However, considering explicitly this model variant, we find that the
  balancing recycling cannot by itself impose a constant threshold.  }

Conversely in scenario 1, the RIDA rate per replication fork is
constant, and one has to rationalize the variation of the binding
affinities.
It seems possible that the binding affinities could change with
growth rate through changes in supercoiling, in similar ways to those
seen in Figure~\ref{fig:fig5} and Additional File 1, 
Figure~\ref{fig:SF3}~\cite{Travers1989,Wang1993}. The levels of
average negative supercoiling are known to increase as the growth rate
increases~\cite{Balke1987}.
\rev{However, it makes sense to challenge the validity of the basic
  assumption that the ratio of DnaA-ATP to DNA at the time of
  initiation is constant. This model assumes that the affinity for
  DnaA-ATP binding to its own promoter can change with growth rate but
  its affinity for the origin does not. The first assumption mainly
  allows the model to change the magnitude of negative autoregulation
  as a function of growth rate, and it may indeed be explained by the
  changes in global cellular parameters such as negative
  supercoiling.  We have considered how the activation threshold in the model (estimated
in Additional File 1, Figure \ref{fig:varying_threshold} and corresponding caption) would be affected if the binding
affinity for DnaA to the origin would vary in the same way as
its value at the dnaA promoter, required by Scenario 1, for a set of
realistic parameters. We found that these changes in $r(X)$ are less
than 10$\%$ over a wide range of growth rates, suggesting that this
scenario might be robust. Indeed, the observed threshold is certainly
approximately constant when compared to the untransformed case i.e.
the different values of the ratio at $t=X$ shown in Figure \ref{fig:fig3}A.}
\rev{More generally, the initiation of DNA replication has been significantly
  simplified in this model; all it requires is a specific amount of
  available DnaA-ATP molecules. However we know that other factors,
  such as the binding of nucleoid proteins FIS, IHF, H-NS and HU, may
  contribute to the formation of an open complex at the origin. On the
  other hand, other recent results have shown that at slower growth
  (slower than the range considered here) the cell contains a greater
  average amount of DnaA-ATP per origin that results in initiation
  events that are independent of the novel synthesis of
  DnaA-ATP~\cite{Flatten2009}. These results suggest that the
  regulation of the initiation process at the origin might indeed be
  dependent on the growth rate and that these changes still remain to
  be characterized quantitatively before they can be included in a
  theoretical model.  }

Interestingly, the basal rate of transcription of the \textit{dnaA}
gene, $k_A$ must vary in both scenarios. Figure~\ref{fig:fig5} shows
that $k_A$ decreases as the cells grow more slowly. This is what is
expected from a promoter like the one of the \textit{dnaA} gene that
closely resembles ribosomal RNA promoters.  This family of promoters
have a GC-rich sequence at the transcription initiation site called a
discriminator region. This region renders the activity of the promoter
sensitive to the degree of negative supercoiling, which activates
transcription by enhancing DNA melting, and leads to its inhibition by
the accumulation of ppGpp at slower growth rates~\cite{Travers2005}.

\section*{Conclusions}
All things considered, we can say that perhaps our main result is that
the determination of the timing of initiation by DnaA, besides relying
on the known ``architecture'' comprising autorepression, RIDA and a
number of other ``dedicated'' processes, can be understood only in its
complex interplay with bacterial physiology (comprising DNA
supercoiling, ppGpp, growth-rate dependent partitioning of molecular
machinery, etc.)

Nevertheless, it makes sense to ask whether this model allows us to
elucidate some features of the reciprocal role of RIDA and DnaA
autorepression, its two main ingredients. 
Biologically, RIDA renders the control of DnaA-ATP dependent upon
ongoing DNA replication, and thus results in an increase in DnaA-ATP
when replication forks are blocked. Autorepression however probably
plays a larger role in the absence of RIDA at slow growth, or in
bacteria that do not have RIDA at all (such as \textit{B.~subtilis}, where DnaA
titration at the replication fork seems to play an important
role)~\cite{Katayama2010}.
The Hda protein and thus the RIDA process seems to be quite specific
to the fast-growing \textit{E. coli} bacterium and its close relatives
in the \textit{Enterobacteriaceae} (UniProtKB), suggesting that in
other bacterial species this level of regulation may not be required
and is replaced instead by protein degradation, e.g. in
\textit{Caulobacter}~\cite{Gorbatyuk2005,Collier2009}, or a high
intrinsic ATPase activity of the protein, as in \textit{Mycobacterium
  tuberculosis}~\cite{Madiraju2006}.
We have verified that the model can work in the absence of
autorepression or RIDA, but the tuning of the parameters to achieve a
constant threshold is more ``difficult'', \rev{in the sense that it
  requires more fine-tuning of the parameters, since the ratio $c_1/P$ should remain
constant with growth rate.
} meaning that it is
possible that a smaller range of growth rates would be accessible in
these conditions.
Moreover, in the model, increasing autorepression or RIDA rate results
in a smaller amplitude of the oscillations of the ratio DnaA-ATP/DNA
during the cell cycle, and in a smaller variation in the average
amount of DnaA-ATP per cell as the growth rate is varied. This may be
advantageous for the use of DnaA as a transcription factor which has
to sense perturbations in the replication status of the cell at all
growth rates.

\section*{Competing Interests}
The authors declare that they have no competing interests.

\section*{Authors Contributions}
MCL, BS, and BB designed research. MG, BB, MCL, UF, and CS performed
research. MG, BS and MCL wrote the paper. All authors read and approved the final manuscript.

\section*{Acknowledgements}
\ifthenelse{\boolean{publ}}{\small}{} 
We are grateful to Matteo Osella, Rosalind Allen, Pietro Cicuta,
Antonio Celani, Andrea Sportiello, Kunihiko Kaneko and Massimo
Vergassola for useful discussions and feedback.  The authors
acknowledge support from the Human Frontier Science Program
Organization (Grant RGY0069/2009-C) and from EPSRC.



{\ifthenelse{\boolean{publ}}{\footnotesize}{\small}
 \bibliographystyle{bmc_article}  
\bibliography{DnaA_bibliography} } 


\ifthenelse{\boolean{publ}}{\end{multicols}}{}


\newpage
\section*{Figures}

\begin{figure}[!ht]
\includegraphics[width=\textwidth]{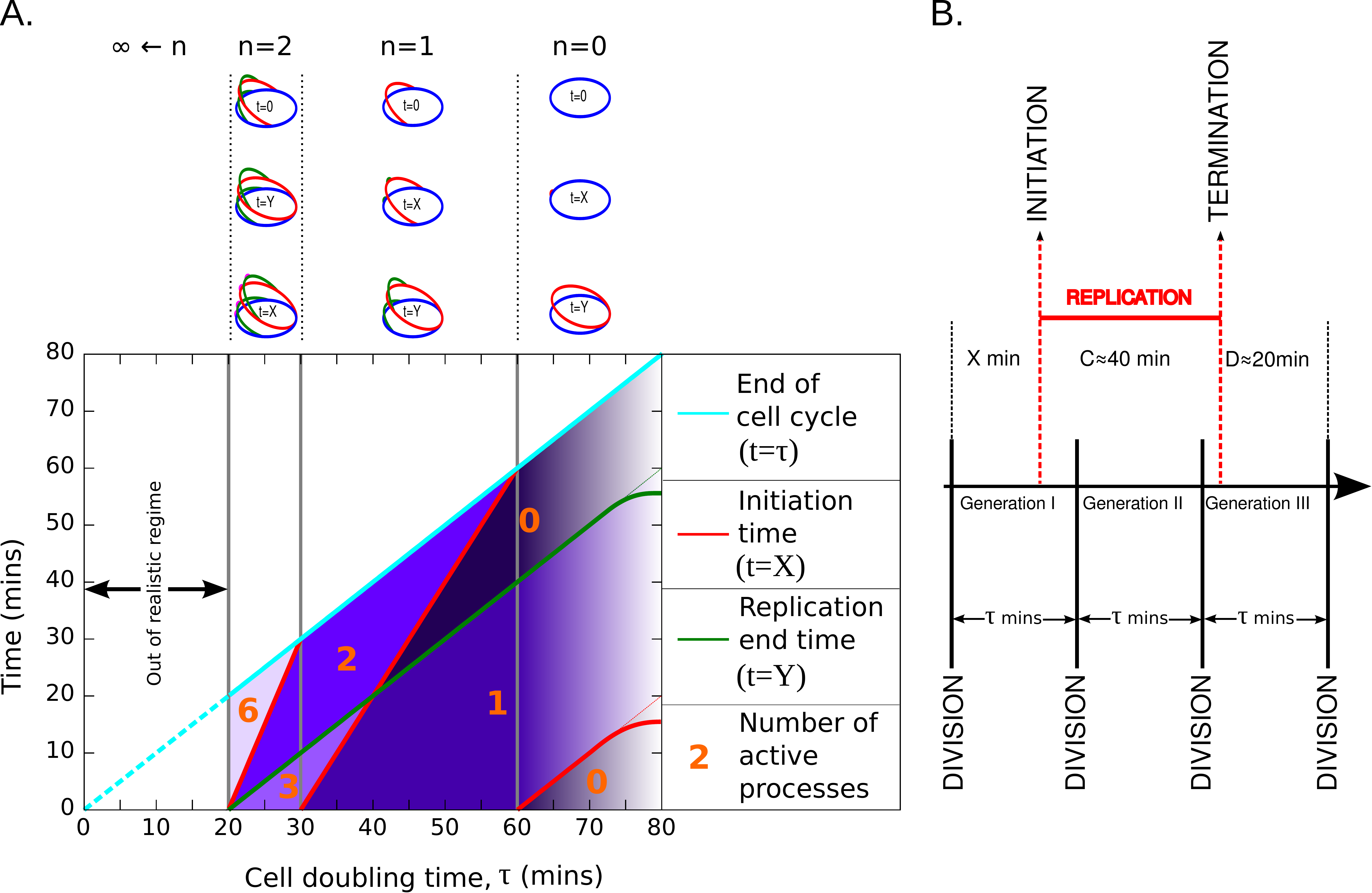}
\caption{\textbf{Timing of DNA replication initiation as a function of
    the length of the cell cycle according to the Cooper and
    Helmstetter model.} A: Plots of the values of $X$, $Y$ and number
  of active processes (termed $\mathcal{F}$ in the main text) in each
  region of the graph, for different values of the cell doubling
  time. The purple shading reflects the number of active processes in
  each region, with lighter shades denoting a greater number of active
  processes. Towards $\tau=80$ mins, the lines $t=X$ and $t=Y$ are
  shown curving off, showing that this is outside the regime $20$ mins
  $\le\tau\le60$ mins where the $C$ and $D$ periods can be considered
  constant. Above the graph in panel A are diagrams of the state of
  the chromosome for critical time values, for each of the values of
  $n$ (the number of overlapping replication rounds). B: Illustration
  of overlapping replication rounds, in the case of a complete
  replication round of $n=2$ overlapping rounds, and $3$
  generations.}  \label{fig:fig1}
\end{figure}

		
\begin{figure}[!ht]
\includegraphics[width=\textwidth]{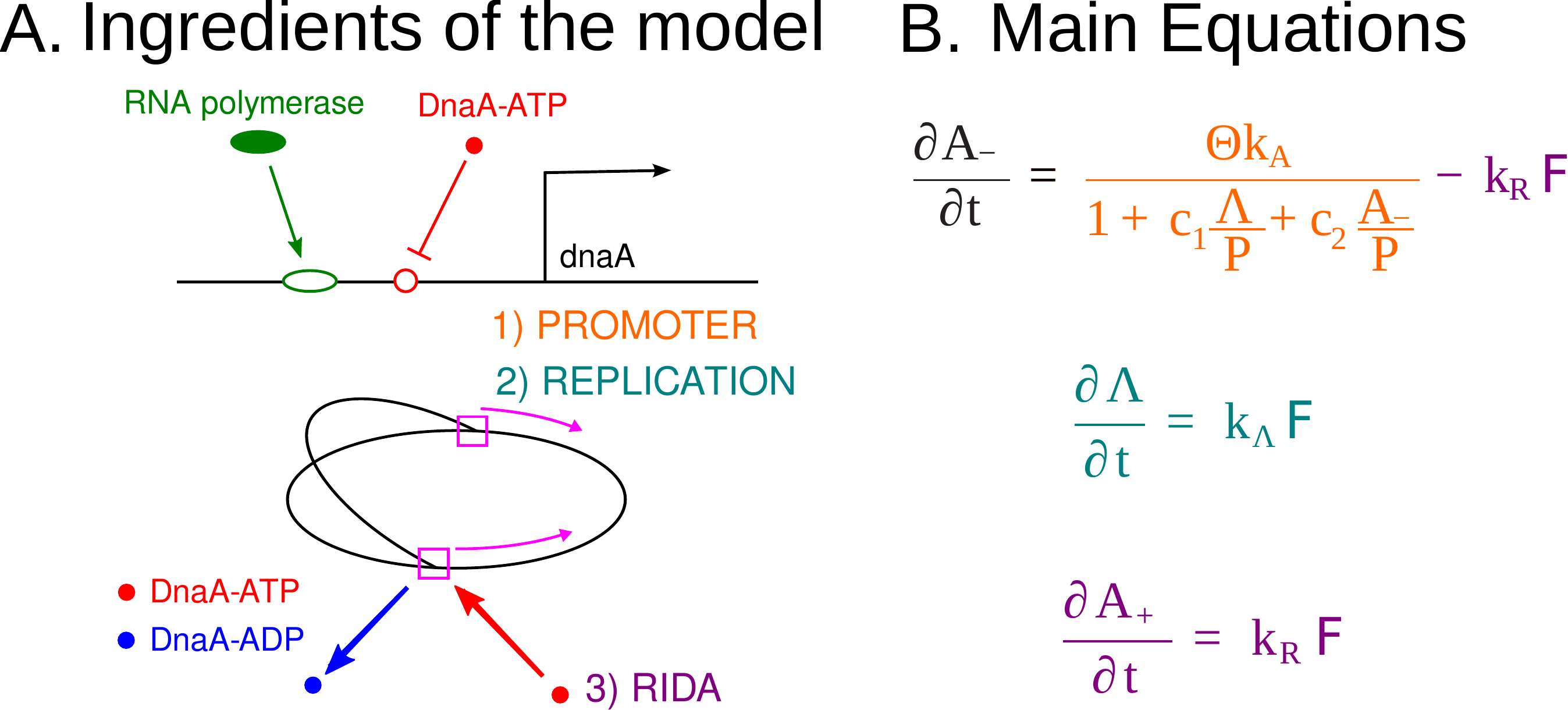}
\caption{\textbf{Ingredients of the model.} A: illustration of 1) the
  autorepression of the $dnaA$ gene 2) the growth of the chromosome by
  the DNA replication process 3) RIDA taking place at the replication
  forks. B: The key equations of the model, with the terms colour
  coded to match the ingredients shown in panel A. The parameters in
  the model are: $A_-$ (number of DnaA-ATP molecules), $A_+$ (number
  of DnaA-ADP molecules), $\Lambda$ (total genome length), $P$ (number
  of RNA polymerase molecules), $\Theta$ (number of $dnaA$ genes),
  $k_A$ (basal transcription rate of one $dnaA$ gene), $c_1$, $c_2$
  (binding constants), $\mathcal{F}$ (number of pairs of replication
  forks), $k_R$ (RIDA rate per replication fork), $k_\Lambda$ (growth
  rate of the chromosome per replication fork). The first equation
  represents the change in the number of DnaA-ATP molecules, with a
  source term due to the $dnaA$ promoter (as all newly synthesised
  DnaA is assumed to bind to ATP due to the relative abundance of ATP
  in the cell), and a sink term due to the conversion of DnaA-ATP to
  DnaA-ADP in RIDA. The second equation represents the growth of the
  genome in the cell. The third equation represents the change in the
  number of DnaA-ADP molecules. The only source term is the same as
  the sink term in the DnaA-ATP equation since it is assumed DnaA-ADP
  is only created from DnaA-ATP during RIDA.}
  \label{fig:fig2}
\end{figure}


\begin{figure}[!ht]
\includegraphics[width=\textwidth]{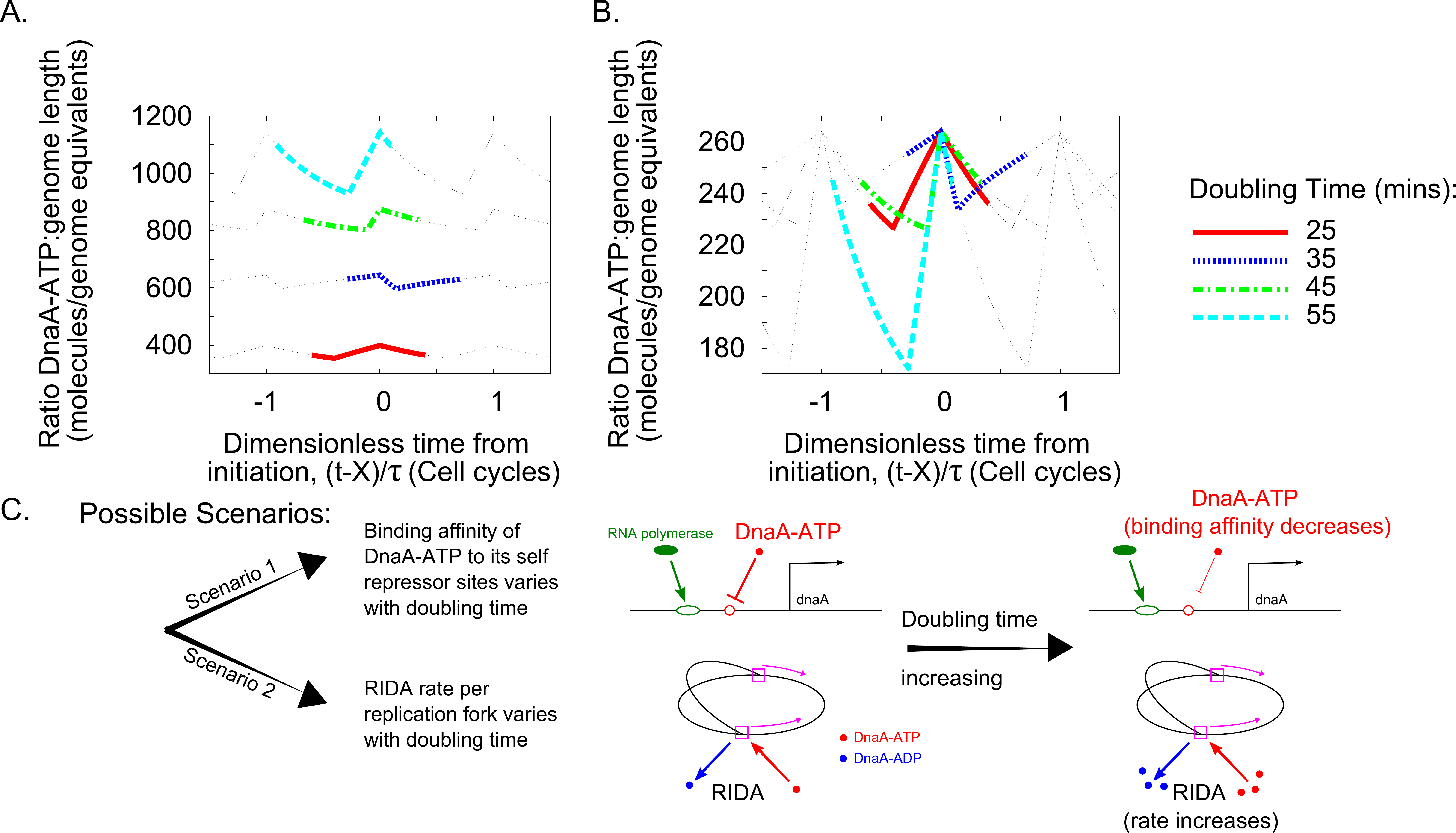}
\caption{\textbf{The model imposes a specific DnaA-ATP threshold at
    the moment of initiation $(t=X)$.} A: The model with fixed
  parameters cannot explain an `initiation threshold' since a
  different value of the ratio DnaA-ATP:genome length ($r$) is
  obtained at initiation ($t=X$) for each value of the cell doubling
  time, $\tau$. B: We perform a mathematical transformation upon the
  model to impose a threshold for the ratio DnaA-ATP:genome length
  ($r$) at the moment of initiation ($t=X$) (in this specific example
  the threshold has been imposed by allowing the binding affinity of
  DnaA-ATP to its self repressor site to vary). In both panels A and
  B, the $x$-axis has been translated and normalized to denote
  fractions of cell cycles with the initiation time given by
  $\frac{t-X}{\tau}=0$. C: In the case in which both autorepression
  and RIDA are included in the model there are two scenarios in which
  an `initiation threshold' can be imposed upon the model. The first
  of these requires the binding affinity of DnaA-ATP to its self
  repressor sites to decrease with increasing cell doubling time.  The
  second scenario requires that the RIDA rate increases with
  increasing cell doubling time. In all scenarios the value of $k_A$
  increases with increasing growth rate.}
  \label{fig:fig3}
\end{figure}

\begin{figure}[!ht]
\begin{center}
\includegraphics[width=0.6\textwidth]{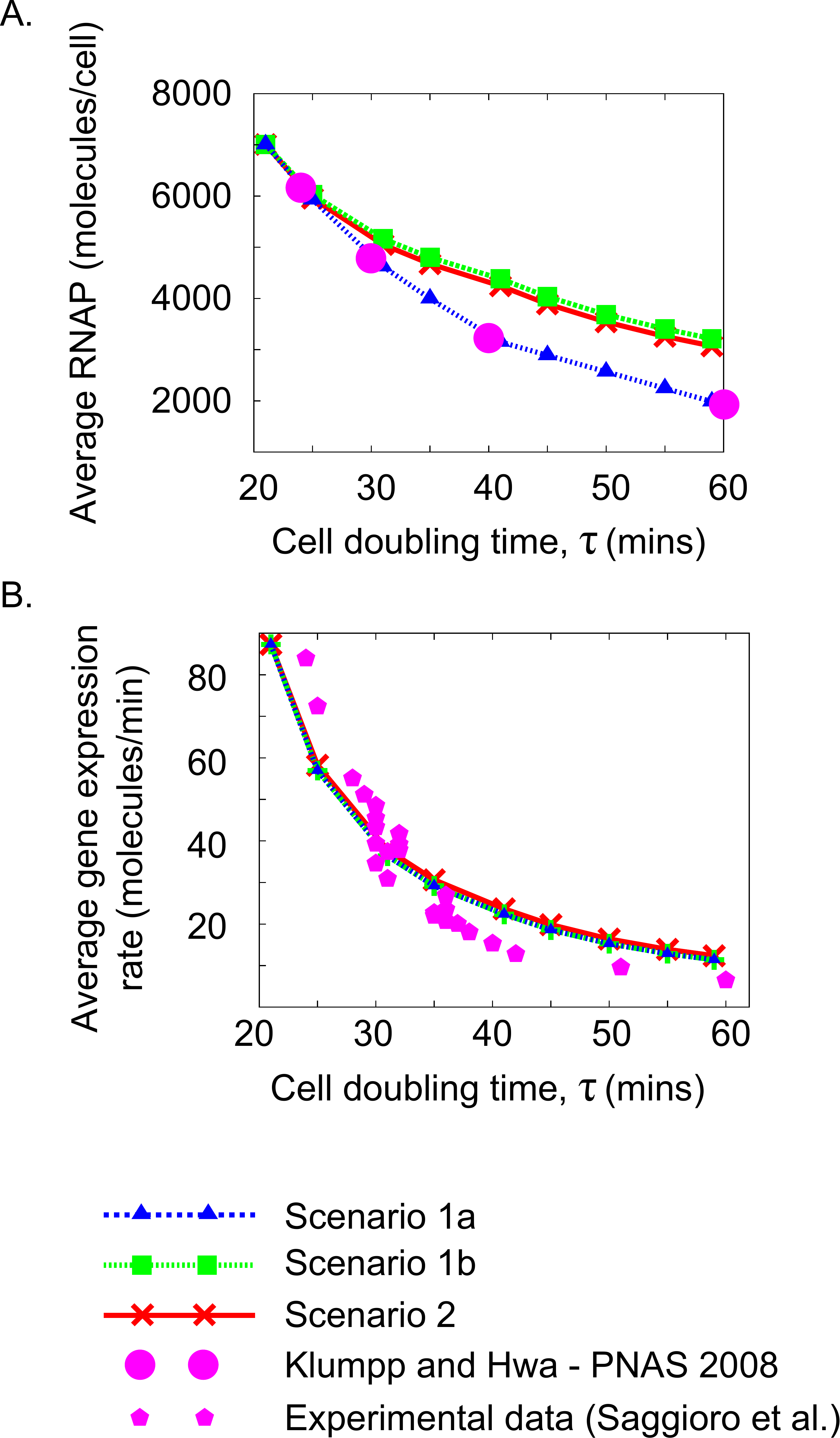}
\end{center}
\caption{\textbf{All scenarios of the model are compatible with
    previous measurements and predictions.}  A: Variation in the
  average number of RNAP molecules per cell with doubling
  time. Simulations from the three scenarios (connected triangles,
  squares and crosses) are compared to the (validated) predictions of
  ref.~\cite{Klumpp2008}. Scenarios 1b and 2 are compatible with the
  results (which are assumed in scenario 1a). B: Variation in the
  average expression rate per cell of the $dnaA$ gene with growth rate
  in the three scenarios (connected triangles, squares and crosses)
  agrees with our direct measurements (Chiara Saggioro, Anne Olliver, Bianca Sclavi: Multiple levels of regulation in the growth rate dependence of DnaA expression, submitted). The
  measurements (pentagons) are obtained with GFP reporters on a
  plasmid, normalized by plasmid number and gene copy-number with
  varying growth rate. \rev{The experiment details are available in
    Additional File 1,~\ref{sec:experimental_methods}}. This
  prediction is also compatible with the results of Chiaramello and Zyskind~\cite{Chiaramello1989}.}
  \label{fig:fig4}		
\end{figure}

\begin{figure}[!ht]
\includegraphics[width=\textwidth]{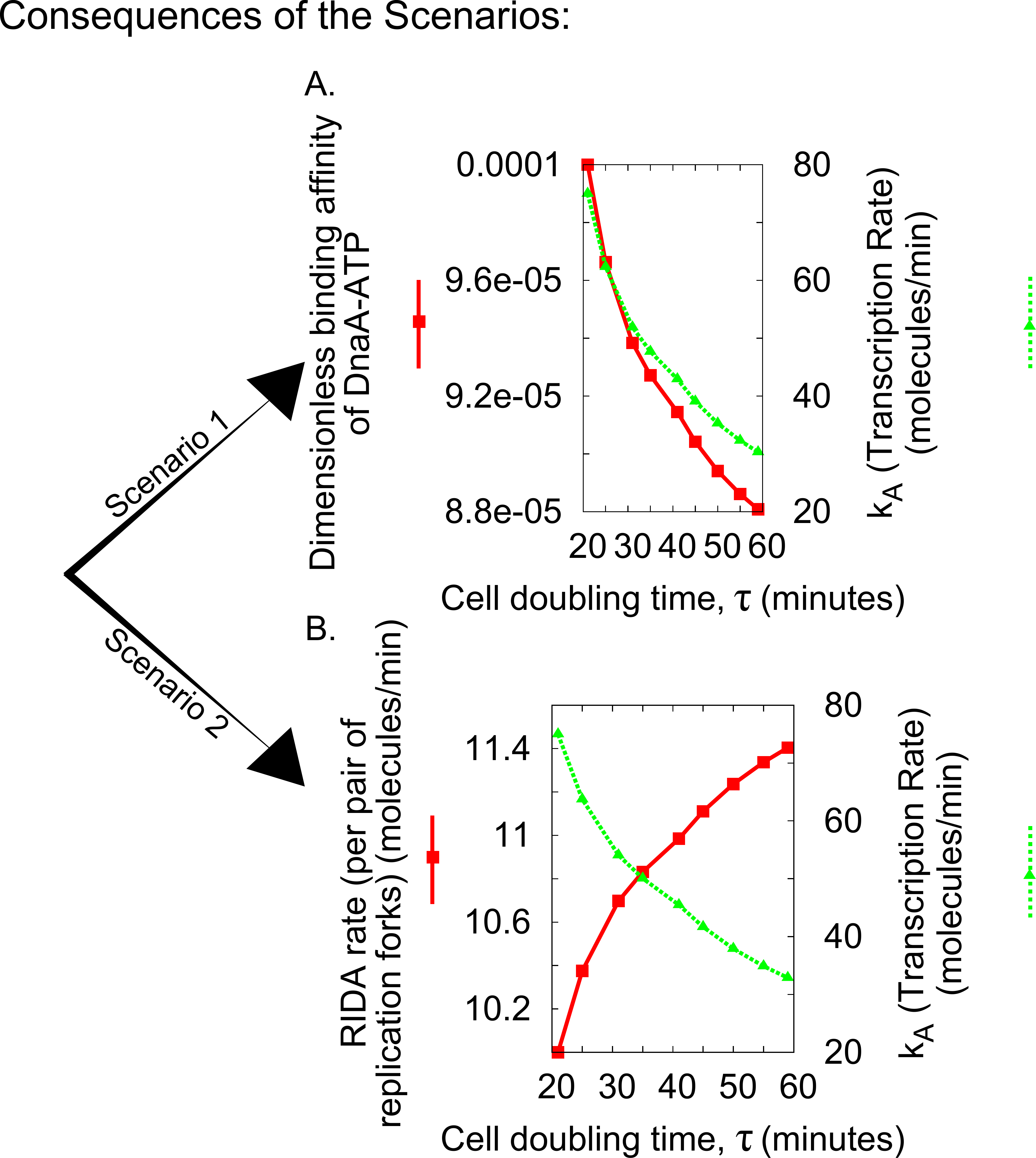}
\caption{\textbf{Predictions of the model can distinguish between
    different scenarios.} Two possible scenarios can result in a
  constant initiation threshold for the model. \rev{In the first the
    binding affinity of DnaA-ATP to its repressor sites decreases with
    increasing cell doubling time, and in the second the RIDA rate
    increases with cell doubling time}. In both scenarios $k_A$ (the
  basal transcription rate) must decrease with increasing cell
  doubling time.}
  \label{fig:fig5}
\end{figure}
%



\clearpage
\section*{Tables}
\begin{table}[!ht]
\captionsetup{singlelinecheck=off}
\caption{\textbf{Initial values for the parameters in the
    model.} } 
  \begin{tabular*}{\textwidth}{ @{\extracolsep{\fill}} l c c c}
    \toprule
    \textbf{Parameter} & \textbf{Untransformed Value} & \textbf{Units} &
    \textbf{Reference} \\
     &  \textbf{(at $\mathbf{\boldsymbol\tau=21}$ mins)} &  & \\
    \midrule
    Basal transcription rate  $k_A$  & 75 & molecules/min &  \cite{Klumpp2008a} \\
    
    RNAP binding $e^{\Delta\epsilon_{pd}/k_BT}$ & 12/10000 & Dimensionless &
    \cite{Bintu2005} \\
    
    Dna-ATP binding $e^{\Delta\epsilon_{ad}/k_BT}$ & 1/10000 & Dimensionless &
    \cite{Speck2001}, $\dag$ \\
    
    RNAP amount  $P_0$ & 5050 & molecules & \cite{Klumpp2008} \\
    
    RIDA rate   $k_R$ & 10 & molecules/min & \cite{Donachie2003} \\
    		
    Replication rate  $k_\Lambda$ & 1/40 & genome equivalents/min & \cite{Cooper1968}
    \\
    	
    non-specific binding sites  $N_{NS}$ & $5\times10^6$ & (genome equivalent)$^{-1}$ &
    \cite{Bintu2005}
    \\	
    \bottomrule
\end{tabular*}
\vskip 0.4cm
\caption*{The parameters are fixed with the values in the table for $\tau=21$ mins and then are able to vary
  for the other values of $\tau$ in order to fix the ratio
  of DnaA-ATP:chromosome length at the moment of initiation
  ($t=X$). $\dag$: Chiara Saggioro, Anne Olliver, Bianca Sclavi: Multiple levels of regulation in the growth rate dependence of DnaA expression, submitted.}
\label{tab:initial_values}
\end{table}

\begin{table}[!ht]
\captionsetup{singlelinecheck=off}
\caption{\textbf{The dependence of the parameters in
    the different scenarios.}}
\begin{center}		
  \begin{tabular*}{\textwidth}{@{\extracolsep{\fill}}c c c }
    \toprule
    \textbf{Scenario} & \textbf{Floating Parameters} & \textbf{Fixed parameters} \\
    \midrule
    1a & $k_A$, $c_1$, $c_2$ & $P$, $k_R$ \\
   
    1b & $k_A$, $c_2$, $P$ & $c_1$ , $k_R$ \\
    
    2 & $k_R$, $k_A$, $P$ & $c_2$, $c_1$ \\
    \bottomrule			
  \end{tabular*}
  \vskip 0.4cm
  \caption*{In all the scenarios, $k_A$ varies with growth rate. In
    scenario 1a, $P$ changes as a function of growth rate with the
    values obtained from ref.~\cite{Klumpp2008}. The dependence of the
    other parameters on growth rate is scenario specific (see
    Additional File 1, Figure~\ref{fig:SF0}).}
  \label{tab:scenarios2}
  \end{center}	
\end{table}

\newpage
\section*{Additional Files}
 \subsection*{Additional file 1 --- Additional Text and Figures}
Single pdf file containing the additional text and figures mentioned
in the main text. 
%

\clearpage
\newpage
\section*{\center{\Large Additional Text and Figures}}

\renewcommand{\thesection}{Section
\arabic{section}}
\setcounter{figure}{0} 
\setcounter{page}{1} 
\setcounter{table}{0} 
\setcounter{section}{0} 
\renewcommand{\figurename}{Figure}
\renewcommand{\thefigure}{A\arabic{figure}}
\renewcommand{\thetable}{A\arabic{table}}
\renewcommand{\refname}{Additional File 1 References}
\vspace{1cm}

\section{Considerations on the initiation mass}
\label{sec:initiation_mass}
The initiation mass of \textit{E. coli}, a concept put forward by
Donachie in 1968 \cite{Donachie1968}, is an idea that, while
independent of the work presented in this paper, has been at the
centre of the research into initiation of chromosome replication since
its inception. This section presents some thoughts on the implications
of our model on the initiation mass argument.

As explained in the main text, steady exponential
growth of the cell mass is assumed. The volume of the
cell at a time $t$, denoted $\Omega(t)$, is thus

\begin{equation}
  \Omega(t)=\Omega_0 e^{\alpha t},
  \label{eq:volume}
\end{equation}
where $\alpha$ is the growth rate such that

\begin{equation}
  \alpha=\frac{\log(2)}{\tau}
  \label{eq:alpha}
\end{equation}

If it is assumed that the initial mass of the cell is an exponential
function of the growth rate \cite{Schaechter1958}, it is possible, for
example, to write

\begin{equation}
  \Omega_0=\Omega_0(\alpha)=\frac{e^{\alpha(C+D)}\Omega_*}{2},
  \label{eq:omega0}
\end{equation}

where $\Omega_*$ is a constant.

With \eqref{eq:volume} and \eqref{eq:alpha}, this implies that

\begin{equation}
  \frac{\Omega(X)}{2^n}=constant=\Omega_*,
  \label{eq:omega*}
\end{equation}
where $2^n$ is the number of \textit{oriC} in the cell. This gives a
constant ratio of \textit{oriC} to mass at initiation and so is in
agreement with what was shown by Donachie \cite{Donachie1968}.

Donachie's observations of a constant initiation mass are based on the
observation by Schaechter et al. \cite{Schaechter1958} of the initial
mass of a cell growing exponentially with growth rate. As discussed,
for the ratio of the mass to number of origins to be constant at
initiation, a very specific initial cell mass is required, namely

\begin{equation}
	\Omega(0)=\frac{\Omega_*}{2}e^{\alpha(C+D)},
\end{equation}

if the Cooper-Helmstetter model is assumed. However, this particular
trend of initial mass with growth rate is not claimed to hold by
Schaechter and coworkers: it is simply an exponential relationship
that is claimed to hold. Thus, if the relationship was slightly
different, say, for example,
$\Omega(0)=\frac{\Omega_*}{2}e^{\alpha(C+D+1)}$ then one would have

\begin{equation}
	\frac{\Omega(X)}{2^n}=\Omega_*e^\alpha,
\end{equation}

i.e. the initiation mass would also be an exponential function of the
growth rate. So, under the assumption of an initial cell mass varying
exponentially with growth rate, it is not necessary to have an initiation mass
that is constant; an initiation mass that is a continuous
function of growth rate is possible. This study is concerned with determining
whether DnaA can fulfill a threshold condition at $t=X$; this question
is independent of whether the origin to mass ratio is constant or not
at $t=X$.

\section{Thermodynamic model for the promoter}
\label{sec:promoter}

This study describes promoter activity using the thermodynamic model
first introduced by Shea and Ackers~\cite{Shea1985,Ackers1982}.  This
section introduces the mathematics of the derivation of this term.
Note that the term derived here is for the simple autorepressor, which
is the form used in the equation~\eqref{eq:full}.

The Shea-Ackers model calculates the probability (assuming equilibrium
binding) that RNAP is bound to a given promoter~\cite{Bintu2005}.  The
validity of the equilibrium binding assumption relies on the rate of
binding and unbinding of RNAP to and from the promoter being much
faster than the rate of open complex formation. The same is true for
the relevant transcription factors (such as DnaA-ATP in this case).
Under these assumptions, the probability that RNAP is bound to a
particular promoter (in the absence of transcription factors) is given
by the ratio of the statistical weight of the state in which RNAP is
bound to that promoter, divided by the sum of the statistical weights
of all possible states.

To formalize this, the number of non-specific binding sites
on the chromosome is denoted as $N_{NS}$ and the number of RNAP molecules as
$P$. The number of ways of arranging all the RNAP molecules in the
non-specific sites is:

\begin{equation}
  \frac{N_{NS}!}{P!(N_{NS}-P)!} \ ,
  \label{eq:arrangements0}
\end{equation}

and the statistical weight of this state is thus

\begin{equation}
Z(P)=\underbrace{\frac{N_{NS}!}{P!(N_{NS}-P)!}}_\text{number of
arrangements} \times
\underbrace{e^{-P\varepsilon^{NS}_{pd}/k_BT}}_\text{Boltzmann
weight (binding energy)}.
\label{eq:stat_weight}
\end{equation}

And so, the probability of having RNAP bound to a given promoter
is given by

\begin{equation}
\mathbb{P}\text{(RNAP
bound)}=\frac{Z(P-1)e^{-\varepsilon^S_{pd}/k_BT}}{Z_{tot}(P)} \ ,
\end{equation}

where $Z_{tot}(P)$ represents the sum of the possible statistical
weights, namely

\begin{equation}
Z_{tot}(P)=\underbrace{Z(P)}_\text{promoter
unoccupied}+\underbrace{Z(P-1)e^{-\varepsilon^S_{pd}/k_BT}}_\text{RNAP
bound to promoter}.
\end{equation}

What can be seen is that this form of the derivation for the
probability of the RNAP being bound to a given promoter does not
require the volume to be considered explicitly. The spatial
distribution of RNAP within the cell is not considered as it is
assumed to provide only a weak perturbation to the probability, given
the large number of RNAP molecules in the cell~\cite{Bintu2005}.

Now consider the probability for the promoter in the both
more relevant and complex case where a repressing transcription factor
is involved (namely DnaA-ATP).

The notation used is as follows: $Q$ is the promoter term; $A$ is the
number of DnaA-ATP molecules as (where the subscript used in the main
text has been dropped for ease of notation); $N_{NS}$ is the number of
non-specific binding sites on the chromosome; $P$ is the number of
RNAP molecules.

The number of ways of arranging the RNAP molecules and the DnaA-ATP
molecules in the non-specific binding sites is
  		
\begin{equation}
  \frac{N_{NS}!}{P!A!(N_{NS}-P-A)!}
  \label{eq:arrangements}
\end{equation}

which is effectively the number of arrangements where the appropriate
promoter is unoccupied. The statistical weight for this can be written

\begin{align}
Z(P,A)=
&\underbrace{\frac{N_{NS}!}{P!A!(N_{NS}-P-A)!}}_\text{Number
of arrangements} \nonumber \\
&\times \underbrace{e^{-P\varepsilon_{pd}^{NS}/k_BT}}_\text{weight
of each RNAP state} \nonumber \\
&\times
\underbrace{e^{-A\varepsilon_{ad}^{NS}/k_BT}}_\text{weight of
    each DnaA-ATP state} \ ,
  \label{eq:Z}
\end{align}
where the exponential terms are the Boltzmann weights.

Thus, under the assumption that DnaA-ATP acts as an autorepressor, the total
statistical weight of all the scenarios is given by

\begin{align}
Z_{tot}(P,A)= &\underbrace{Z(P,A)}_\text{empty sites}
\nonumber \\
&+\underbrace{Z(P-1,A)e^{-\varepsilon_{pd}^{S}/k_BT}}_\text{RNAP
on promoter} \nonumber \\
&+\underbrace{Z(P,A-1)e^{-\varepsilon_{ad}^{S}/k_BT}}_\text{DnaA-ATP
on specific site}  \ .
\end{align}

The probability that RNAP is bound at the \textit{dnaA} promoter is
\begin{align}
\mathbb{P}(\text{RNAP bound}) &=
\frac{Z(P-1,A)e^{-\varepsilon_{pd}^{S}/k_BT}}{Z_{tot}(P,A)}
\nonumber \\
&= \frac{1}{1+\frac{N_{NS}}{P}e^{\Delta
\varepsilon_{pd}/k_BT}+\frac{A}{P}e^{(\Delta \varepsilon_{pd}-\Delta
\varepsilon_{ad})/k_BT}},
  \label{eq:promoter}
\end{align}	

where it has been assumed that $N_{NS}\gg P$ and $A$.

The basal rate of transcription of the \textit{dnaA} promoter is then defined as $k_A$ and the copy-number of
\textit{dnaA} promoters in a given cell at one time (which can be
computed from the Cooper-Helmstetter model) as $\Theta(t,\tau)$.

The same kind of argument can be used in order to justify the fact
that one expects a constant initiation threshold for DnaA.
Consider the fact that $20$ DnaA molecules must bind to the origin
for initiation to occur. It can be assumed that this state is binary,
i.e. that either 20 molecules are bound, or 20 molecules are not
bound. The probability of having 20 molecules bound is thus:

\[\mathbb{P}(\text{Origin
  Bound})=\frac{Z(A-20)e^{-20\varepsilon_{ao}/k_BT}\omega}{Z(A-20)e^{-20\varepsilon_{ao}/k_BT}\omega+Z(A)},\]

where
\[Z(A)=\frac{N_{NS}!}{A!(N_{NS}-A)!}.\]

Dividing through, and under the assumptions above, gives that

\begin{equation}
\mathbb{P}(\text{Origin Bound})=\frac{1}{1+\frac{e^{20\Delta
      \varepsilon_{ao}/k_BT}}{\omega(\kappa r)^{20}}}.
      \label{eq:prob_origin_bound}
\end{equation}

Thus, there is a value of $r$, determined by $\frac{e^{20\Delta
    \varepsilon_{ao}/k_BT}}{\omega\kappa ^{20}}$ at which the
probability rapidly approaches unity. This can be seen as the value of
$r$ where initiation takes place. Note that this estimate is rather
robust to the addition of other possible states, such as states in
which DnaA-ATP can be bound to other sites, assuming that the number
of these extra specific sites is small in comparison to the total pool
of DnaA-ATP molecules.

\subsection{Considerations on the increase in RNAP levels}
\label{sec:RNAP_levels}
A further factor to consider is the way in which RNAP levels vary
during the cell cycle. As with other elements of the cell, the number
of RNAP molecules must double during each cell cycle when the bacteria
are in exponential growth. The assumption made here is that the RNAP
levels grow exponentially during the cell cycle, to reach double their
initial value at the end of the cell cycle. Combined with the
assumption that the volume of the cells grows exponentially, this
means that the concentration of RNAP stays constant through the cell
cycle, in agreement with the results of Oeschger et
al.~\cite{Oeschger1974}. 
Moreover, even if the levels of RNAP grow
linearly, this is only a weak perturbation from the exponential
growth. Consequently, the expectation is that this will not
significantly affect the model dynamics.

\section{Parameter transformation enforcing a constant threshold} 
\label{sec:math_trans}

This section discusses the parameter transformation of equation
\eqref{eq:full}, providing the key to probing how the parameters of
the model must vary with growth rate. As discussed in the main text,
equation \eqref{eq:full} is transformed to fix the value of $r$ at the
moment of initiation ($t=X$) to be the same for every value of $\tau$
in the range $20<\tau<60$. This is enforced by a translation and
scaling on $r$ such that

\begin{equation}
  r'(X,\tau)=\lambda(\tau)r(X,\tau)-a(\tau) \ ,
  \label{eq:transformation}
\end{equation}

where $r'$ is the translated version of $r$.\pb
The wish is to see what must happen to the coefficients in the equation for the fixed
threshold condition to hold, when the structure of \eqref{eq:full} is preserved in the transformation. The same transformation can be applied in
numerical simulations, but its analytical form is more instructive
with respect to classifying the possible parameter
transformations. \pb

Equation \eqref{eq:full} can be rewritten in the desired form, with translated
variables, as

\begin{equation}
  \frac{\partial r'}{\partial t} = \frac{1}{\Lambda}
  \left(
 \frac{\Theta k'_A}
      {1+c'_1 \frac{\Lambda}{P'} + c'_2 \frac{\Lambda}{P'}r'}
      -  (k'_R+r'k_\Lambda) \mathcal{F}
 \right) \ .  
 \label{eq:fulltransformed}
\end{equation}

Note that $k_\Lambda$ cannot change, as this rate is externally
imposed by the Cooper-Helmstetter model.\pb
The substitution of $r$ from \eqref{eq:transformation} into equation
\eqref{eq:full} gives

\begin{equation}
\frac{\partial r'}{\partial t}=\lambda \frac{\partial
r}{\partial t}=\frac{\lambda}{\Lambda}\left(\frac{\Theta
k_A}{1+c_1
\frac{\Lambda}{P}+c_2\frac{\Lambda}{P}(\frac{r'+a}{\lambda})}-(k_R+\left(\frac{r'+a}{\lambda}\right)k_\Lambda)\mathcal{F}\right).
\label{eq:fulltransback}
\end{equation}

Comparison of the coefficients of $r'$ and $\Lambda$ in equation
\eqref{eq:fulltransback} with equation \eqref{eq:fulltransformed},
yields the conditions

\begin{equation}
k_A'=\lambda k_A,
\label{eq:case0.1}
\end{equation}

\begin{equation}
  \lambda k_R + a k_{\Lambda} = k'_R,
\label{eq:kBdash}
\end{equation}

\begin{equation}
\frac{c_2}{P\lambda}=\frac{c_2'}{P'},
\label{eq:c2}
\end{equation}

\begin{equation}
\frac{1}{P}\left(c_1+c_2\frac{a}{\lambda}\right)=\frac{c_1'}{P'},
\label{eq:c1andc2}
\end{equation}

and the condition that defines the transformation, namely

\begin{equation}
r'(X)=\lambda r(X) - a= constant.
\label{eq:threshold}
\end{equation}

At this point we note the assumption that the concentration of RNAP is
constant throughout the cell cycle. Thus
\begin{equation}
P=P_0e^{\alpha t}
\end{equation}

and 
\begin{equation}
P'=P_0'e^{\alpha t}
\end{equation}

Thus, since equations \eqref{eq:c2} and \eqref{eq:c1andc2} have
$e^{\alpha t}$ as a common factor, it is possible to divide through by it.  From
hereafter in this section, $P$ and $P'$ actually refer to $P_0$ and
$P_0'$ but the subscripts are dropped for ease of notation.

Furthermore, equations \eqref{eq:c2} and \eqref{eq:c1andc2} can
combine to give

\begin{equation}
	a=\frac{c_1'}{c_2'}-\lambda\frac{c_1}{c_2}
\label{eq:c1andc2dash}
\end{equation}

Examination of equation \eqref{eq:binders}, reveals that
\begin{align}
 c_1 &= \frac{b_0}{\kappa} \nonumber \\
 c_2 &= \frac{b_0}{b_1}
\end{align}

where $b_0$ and $b_1$ are given by equation \eqref{eq:binders}.  Thus,
writing $c_1'=\frac{b_0'}{\kappa}$ and $c_2'=\frac{b_0'}{b_1'}$ one
sees that equation \eqref{eq:c1andc2dash} can be rewritten

\begin{equation}
	a=\frac{1}{\kappa}(b_1'-\lambda b_1)
	\label{eq:binderdash}
\end{equation}
 
 where $b_1=e^{\Delta \varepsilon_{ad}/k_BT}$.
 
 Now, mathematically, two things can be chosen to be fixed to satisfy
 these equations

\begin{enumerate}
\item The RIDA rate can be fixed: $k_R=k_R'$. It then follows from
\eqref{eq:kBdash} that
\begin{equation}
	a=\frac{k_R}{k_\Lambda}(1-\lambda),
	\label{eq:a1}
\end{equation}

which, when combined with equation \eqref{eq:binderdash} gives

\begin{equation}
 b_1'=\frac{\kappa k_R}{k_\Lambda}(1-\lambda)+\lambda b_1.
 \label{eq:b1dash}
 \end{equation}

 Equation \eqref{eq:a1}, when taken with \eqref{eq:threshold}, gives
 the value that $\lambda$ must take, namely

\begin{equation}
\lambda=\frac{r'(X)+\frac{k_R}{k_\Lambda}}{r(X)+\frac{k_R}{k_\Lambda}}.
	\label{eq:lambda1}
\end{equation}

Either $c_1$ (i.e. $b_0$) or $P$ can then be fixed. 
\begin{enumerate}
\item Since it is possible to choose $c_1'$, it is logical to set
  $c_1'=c_1$. This gives that, $c_2'=\frac{b_0}{b_1'}$ and so
  
  \begin{equation}
P'=\frac{b_1 P \lambda}{\left(\kappa \frac{k_R}{k_\Lambda}
(1-\lambda) + \lambda b_1\right)}.
\end{equation}
  
Note that any trend for $b_0$ could have been chosen. The decision to set
$b_0'=b_0$ here, was due to it being the simplest logical choice. This equation, taken
with equations \eqref{eq:case0.1}, \eqref{eq:b1dash} and
\eqref{eq:lambda1} determines the transformation.
  
\item It is possible to fix $P'$ to a known trend e.g. the one given
  in ref.~\cite{Klumpp2008}. This will then have
  implications for how $c_1$ (and hence $b_0$) must vary. Given that
  $P'$ is now a known quantity, this gives that
          
  \begin{equation}
    b_0'=\frac{b_0 P' \left(\kappa
        \frac{k_R}{k_\Lambda}(1-\lambda)+\lambda b_1\right)}{b_1 P
      \lambda}.
  \end{equation}
  
  This equation, taken with equations \eqref{eq:case0.1},
  \eqref{eq:b1dash} and \eqref{eq:lambda1} determines the
  transformation.
\end{enumerate}
\item It is possible to set $c_1=c_1'$ and $c_2=c_2'$. From this it
follows that
  
  \begin{equation}
    a=\frac{c_1}{c_2}(1-\lambda)
    \label{eq:case2.1}
  \end{equation}
  
  and 
  
  \begin{equation}
    P'=\lambda P.
    \label{eq:case2.2}
  \end{equation}
  
  Furthermore,
  
  \begin{equation}
k_R'=\frac{k_R-\frac{c_1}{c_2}(1-\lambda)k_\lambda}{\lambda}.    \label{eq:case2.3}
  \end{equation}
  
  Thus, equations \eqref{eq:case2.1}, \eqref{eq:case2.2} and
  \eqref{eq:case2.3}, along with
  \begin{equation}
    \lambda=\frac{r'(X)+\frac{c_1}{c_2}}{r(X)+\frac{c_1}{c_2}}
    \label{eq:case2.4}
  \end{equation}
  
  and \eqref{eq:case0.1}, determine the transformation, when $c_1$
  and $c_2$ are fixed.
\end{enumerate}


\section{Average over cell population}
\label{sec:average}

This section discusses how to compute average levels of RNA polymerase
and average gene expression, used for example in
Figure~\ref{fig:fig4}. These average levels are quantities that can be
in principle measured directly.

Consider, for example, the expression level of the \textit{dnaA}
promoter (i.e. $Q$) (presented in Figure~\ref{fig:fig4}B). Denote
the average colony expression level as:

\begin{equation}
	\text{Average colony expression level } = \langle Q(\alpha)\rangle
\end{equation}

where steady exponential growth is assumed and $\alpha$ is the growth
rate. In order to evaluate this average, the probability of finding a
cell at a time $t \in (0,\tau)$ of the cell cycle is required.

Given the exponentially growing colony with growth rate $\alpha$, let
$N(t)$ denote the number of bacteria in the colony at time $t$.  If
the interval $(0,\tau)$ is split into $\tau/\delta t$ equal size
segments each of infinitesimal size $\delta t$, then the number of
bacteria born in the infinitesimal time inverval $(t, t+\delta t)$ is

\begin{equation}
	\delta N(t) = \frac{\ud N}{\ud t}\delta t \ ,
\end{equation}

i.e.

\begin{equation}
	\delta N(t) = \alpha N \ud t=\alpha N_0 e^{\alpha t}\delta t.
\end{equation}

Within a population of bacteria at time $t'$, the probability of
finding a bacterium born in the time interval $(t_b, t_b+\delta t)$
(with $t_b<t'$) is given by

\begin{equation}
  \frac{\delta N(t_b)}{N(t')}=\frac{\alpha N_0 e^{\alpha t_b}}{N(t')}\delta t=\alpha e^{\alpha(t_b-t')} \delta t.
\end{equation}

This is equivalent to finding a bacterium with an age in the range
$(t, t+\delta t)$ where $t=t'-t_b$, and so it can be written that the
probability of finding a bacterium with an age in the range $(t,
t+\delta t)$ is

\begin{equation}
	\mathcal{P}(t, \alpha)=\alpha e^{-\alpha t}\delta t.
\end{equation}

Moreover, since bacteria divide after a time $\tau$, the probability
of finding a bacterium at a stage in the range $(t, t+\delta t)$ of
the cell cycle is given by

\begin{align}
\mathbb{P}(t, \alpha)&=\sum_{n=0}^\infty \mathcal{P}(n\tau+t) \nonumber \\
&=\alpha e^{-\alpha t}\sum_0^\infty e^{-\alpha n \tau} \delta t \nonumber
\\
&=\alpha e^{-\alpha t}\left(\frac{1}{1-e^{-\alpha \tau}}\right) \delta t
\nonumber \\
&= \alpha e^{-\alpha t}\left(\frac{1}{1-\frac{1}{2}}\right) \delta t
\nonumber \\
			&= 2 \alpha e^{-\alpha t} \delta t \nonumber \\
			&= \phi (t,\alpha) \delta t
\end{align}

Note that the individual bacteria are assumed to grow exponentially
with growth rate $\alpha$ in the same way as the colony itself.  Thus,

\begin{align}
  \langle Q(\alpha)\rangle 	& = \sum_{m=0}^{\tau/\delta t} Q(m \delta t,\alpha)\mathbb{P}(m\delta t,\alpha) &\nonumber \\
  & = \sum_{m=0}^{\tau/\delta t} Q(m \delta t,\alpha)\phi(m\delta t,\alpha) \delta t  &\nonumber \\
  & = \int_0^\tau Q(t, \alpha) \phi(t, \alpha) \ud t &\text{ in the
    limit as } \delta t \rightarrow 0
\end{align}

This argument gives the average value of any observable quantity. For
example to get the average of $P$, as in Figure~\ref{fig:fig4}A, one would take

\begin{equation}
  \langle P(\alpha)\rangle =  \int_0^\tau P(t, \alpha) \phi(t, \alpha) \ud t .
\end{equation}

\section{Considerations on other forms of the promoter}
\label{sec:complex_promoter}

This section deals with thermodynamic models for promoters that
incorporate some further experimental findings on the behaviour of the
$dnaA$ gene~\cite{Donachie2003}. Data from footprint experiments has
provided an insight into the cooperativity of the DnaA boxes on the
$dnaA$ gene, which has been proposed to be important for the correct
timing of DNA replication in \textit{Escherichia coli}, as well as
potential autoactivation of the gene~\cite{Speck1999}. 
The models
considered are (i) a promoter with no autorepression, (ii) a model
with cooperativity of DnaA binding to the repression sites at the
\textit{dnaA} promoter, (iii) a promoter with both cooperativity and
autoactivation. The reasons for not choosing these model variants as
the main model formulation are then discussed.

\subsection{Promoter with no autorepression}
\label{sec:noAR}
This section considers a simpler model for the promoter than the one
used in the main text, namely one that has no autorepression. The form
for this type of promoter was derived in~\ref{sec:promoter}
and its implications are discussed more fully here.  When the results
from~\ref{sec:promoter} are used, the main equation takes the
form

\begin{equation}
\frac{\partial r}{\partial
t}=\frac{1}{\Lambda}\left(\frac{\Theta
k_A}{1+c_1\frac{\Lambda}{P}}-(k_R+rk_\Lambda)\mathcal{F}\right).
\end{equation}

This is integrated, with fixed parameters $k_A$, $c_1$, $k_R$ and
$k_\Lambda$ assumed, to give $r(t,\tau)$, for a given $\tau$. The
parameter $c_1$ is then varied (for a fixed value of $\tau$) to probe
the effect that autorepression has (see Figure~\ref{fig:SF5}).  What
can be see in Figure~\ref{fig:SF5} is that autorepression causes the
range of values for $r(t,\tau)$ (the oscillation size), to be much
narrower than for the non-autorepressed case. This might be an
important role that autorepression plays in fine tuning the timing of
replication initiation, since it ensures that, even before any other
parameters are considered to be variable, the differences in the ratio
DnaA-ATP:DNA length is reduced across different $\tau$. For this
reason, autorepression was chosen to be included in the main model
formulation. 


\subsection{Promoter with cooperativity of DnaA binding}
This section discusses a model for a promoter in which there are two
binding sites for DnaA. First, it deals with a form for a simple case
of cooperativity, and then it introduces a more complex form including
differential action of DnaA-ATP and DnaA-ADP.

\subsubsection{Simple cooperativity}
In the model considered here, there are two binding sites for DnaA,
both of which have a repression effect when bound, and which bind
highly cooperatively. The sites are considered to have a low affinity
for DnaA so only a probability to the situation in which both sites
are bound is allocated .  As in \ref{sec:promoter}, the number of
DnaA-ATP molecules is denoted as $A$, the number of non-specific
binding sites as $N_{NS}$ and the number of RNA polymerase molecules
as $P$. Furthermore, the binding energy between a molecule $x$ and a
(non-)specific site on the DNA is defined as $\varepsilon_{xd}^{(N)S}$
and the difference is denoted as
\begin{equation}
\Delta \varepsilon_{xd}=\varepsilon_{xd}^{S}-\varepsilon_{xd}^{NS}.
\end{equation}

In addition, the cooperativity of binding between two DnaA-ATP
molecules is written as $\omega$. 

Thus, the sum of all the statistical weights is
\begin{equation}
Z_{tot}=\underbrace{Z(P,A)}_\text{Empty
sites}+\underbrace{Z(P-1,A)e^{-\varepsilon_{pd}^{S}}}_\text{RNAP
bound}+\underbrace{Z(P,A-2)e^{-2\varepsilon_{ad}^S}\omega}_\text{DnaA-ATP
bound cooperatively} \ .
\label{eq:coop}
\end{equation}
This form allocates zero probability that DnaA-ATP can bind by itself.
This leads directly to the probability of RNAP being bound to the promoter,
namely

\begin{equation}
\mathbb{P}\text{(RNAP
bound)}=\frac{Z(P-1,A)e^{-\varepsilon_{pd}^{S}}}{Z_{tot}} \ ,
\end{equation}

i.e.

\begin{equation}
\mathbb{P}(\text{RNAP bound})=\frac{1}{1+e^{\Delta
\varepsilon_{pd}}\frac{N_{NS}}{PF_{reg}}} \ ,
\end{equation}

where

\begin{equation}
F_{reg}^{-1}=1+e^{-2\Delta
\varepsilon_{ad}}\omega\left(\frac{A}{N_{NS}}\right)^2.
\end{equation}

Figure~\ref{fig:SF4}C shows that the presence of
cooperativity alone cannot produce a constant threshold. Thus, it is
necessary to try and impose a constant threshold on the model using a
transformation such as that described in \ref{sec:math_trans}. Now,
if it is written that $r\kappa=\frac{A}{N_{NS}}$ (see main text for definition of
$\kappa$), an attempt to perform the same transformation as
in~\ref{sec:math_trans}, i.e.
\begin{equation}
r'=\lambda r-a \ ,
\end{equation}
shows that the $(r\kappa)^2$ term produces a linear term in
$r'$ that is not present in the original form of $F_{reg}$ and so the
transformation will fail as it will not generally be possible to
retain the original form of the model.

\subsubsection{A more sophisticated model for cooperativity}
We consider here the more realistic situation where the two binding
sites differ. One is a high-affinity site, which does not contribute
towards autorepression, and the other is a low-affinity site to which
DnaA binds cooperatively and such that, when bound, DnaA has a
repression effect on the \textit{dnaA} gene~\cite{Speck2001}.  The
number of ways of distributing $P$ RNA polymerase molecules and $A$
DnaA-ATP molecules in $N_{NS}$ binding sites is thus

\begin{equation}
Z(P,A)=\frac{N_{NS}!}{P!A!(N_{NS}-P-A)}\times
e^{-P\varepsilon_{pd}^{NS}}e^{-A\varepsilon_{ad}^{N_{NS}}} \ .
\end{equation}

There are five different possible states of the system, described in
the sum of the statistical weights:

\begin{align}
Z_{tot} 	& = \underbrace{Z(P,A)}_\text{Empty sites} \nonumber \\
		& + \underbrace{Z(P-1,A)e^{-\varepsilon_{pd}^{S}}}_\text{RNAP bound} \nonumber \\
		& + \underbrace{Z(P-1,A-1)e^{-(\varepsilon_{pd}^{S}+\varepsilon_{ad}^{H})}}_\text{RNAP bound and high affinity site bound} \nonumber \\
		& + \underbrace{Z(P,A-1)e^{-\varepsilon_{ad}^{H}}}_\text{High affinity site bound and RNAP not bound} \nonumber \\
		& + \underbrace{Z(P,A-2)e^{-(\varepsilon_{ad}^{H}+\varepsilon_{ad}^{L})}\omega}_\text{DnaA-ATP bound cooperatively} \ .
\label{eq:coop2}
\end{align}

This model allocates zero probability that DnaA-ATP can bind to the
low-affinity site alone. For it to bind to this site it must do so
cooperatively once the high affinity site is bound.

This leads to the probability of RNAP being bound to the promoter,
namely:

\begin{equation}
\mathbb{P}\text{(RNAP
bound)}=\frac{Z(P-1,A)e^{-\varepsilon_{pd}^{S}}+Z(P-1,A-1)e^{-(\varepsilon_{pd}^{S}+\varepsilon_{ad}^{H})}}{Z_{tot}} \ ,
\end{equation}

i.e.

\begin{equation}
\mathbb{P}(\text{RNAP bound})=\frac{1}{1+e^{\Delta
\varepsilon_{pd}}\frac{N_{NS}}{PF_{reg}}}
\end{equation}

where

\begin{equation}
F_{reg}^{-1} = 	\frac{1 + \frac{A}{N_{NS}}e^{-\Delta \varepsilon_{ad}^H} + \omega\left(\frac{A}{N_{NS}}\right)^2e^{-(\Delta \varepsilon_{ad}^H+\Delta \varepsilon_{ad}^L)}}
			{1 + \frac{A}{N_{NS}}e^{-\Delta \varepsilon_{ad}^H}}.
\end{equation}

Now, writing $r\kappa=\frac{A}{N_{NS}}$ (see main text for definition
of $\kappa$), the same transformation as for the main
equation can be performed, following the steps in~\ref{sec:math_trans}, i.e.  writing

\begin{equation}
r'=\lambda r-a
\end{equation}

and collecting and comparing terms in $r'$.

\subsection{Promoter with both cooperativity and autoactivation}
This section considers an even more realistic form for the promoter
which incorporates both autoactivation and cooperativity of DnaA
binding of both forms of bound DnaA. 
The main conclusion is that this representation introduces too many
new parameters in the model, which cannot be estimated from
experiments, and therefore adds uncontrolled uncertainty. Hence, priority was given to the simpler and more controlled model presented in
the main text.


First, consider the $dnaA$ gene itself, which has in fact two
promoters, $dnaAp1$ and $dnaAp2$. Since the expression of the $dnaA$
gene is controlled mostly by the $dnaAp2$ promoter, only this
promoter is considered. There are 5 DnaA boxes on the $dnaA$ gene, two high
affinity boxes (1 and 2) which bind both DnaA-ATP and DnaA-ADP and
three low affinity boxes (a, b and c) which bind only DnaA-ATP. Since
box a appears to only affect the $dnaAp1$ promoter, it is ignored for this derivation. DnaA boxes b and c appear to cause autorepression when
bound by DnaA-ATP. Boxes 1 and 2 appear to cause autoactivation when
bound by either DnaA-ATP or DnaA-ADP in situations when DnaA boxes b
and c are unbound.  Moreover, it is assumed that, due to cooperativity,
there must be pairwise binding between boxes 1 and 2 and boxes b and c
i.e. if box 1 is bound by DnaA-ATP then box 2 must also be bound by
DnaA-ATP ~\cite{Speck1999}.

In a similar manner to the previous section, the number of
DnaA-ATP molecules is denoted as $A_-$, the number of DnaA-ADP molecules as
$A_+$, the number of non-specific binding sites as $N_{NS}$ and the
number of RNA polymerase molecules as $P$. Furthermore, the
binding energy between a molecule $x$ and a
(non-)specific site, $i$, on the DNA is defined as $\varepsilon_{xd}^{(N)S_i}$ and
the difference as

\begin{equation}
\Delta \varepsilon_{xd}=\varepsilon_{xd}^{S_i}-\varepsilon_{xd}^{NS} \ .
\end{equation}

In addition to this, the binding energy (cooperativity) between two molecules $a$ and $b$ is written as $\varepsilon_{ab}$. So, the
statistical weight of distributing $P$ RNA polymerase molecules, $A_-$
DnaA-ATP molecules and $A_+$ DnaA-ADP molecules among $N_{NS}$
non-specific binding sites is

\begin{equation}
Z(P,A_-,A_+)=\frac{N_{NS}!}{P!A_-!A_+!(N_{NS}-P-A_--A_+)}\times
e^{-P\varepsilon_{Pd}^{{NS}}}e^{-A_-\varepsilon_{A_-d}^{{NS}}}e^{-A_+\varepsilon_{A_+d}^{{NS}}} \ .
\end{equation}

The sum of all the statistical weights is

\begin{align}
Z_{tot}&=Z(P,A_-,A_+)+Z(P-1,A_-,A_+)e^{-\varepsilon_{Pd}^{S}}
\nonumber \\
&+Z(P,A_--2,A_+)k_{b,c}+Z(P-1,A_--2,A_+)e^{-\varepsilon_{Pd}^{S}}k_{1,2}^-\omega_{0,1}^-
\nonumber \\
&+Z(P,A_--2,A_+)k_{1,2}^-+Z(P,A_-,A_+-2)k_{1,2}^+ \nonumber \\
&+Z(P,A_--4,A_+)k_{b,c}k_{1,2}^-\omega_{b,1}^-+Z(P-1,A_-,A_+-2)e^{-\varepsilon_{Pd}^{S}}k_{1,2}^+\omega_{0,1}^+
\nonumber \\
&+Z(P,A_--2,A_+-2)k_{b,c}k_{1,2}^+\omega_{b,1}^+,
\end{align}

where

\begin{align}
k_{b,c}&=e^{-2\varepsilon_{A_-d}^{S_{b,c}}} \nonumber \\
k_{1,2}^-&=e^{-2\varepsilon_{A_-d}^{S_{1,2}}} \nonumber \\
k_{1,2}^+&=e^{-2\varepsilon_{A_+d}^{S_{1,2}}} \nonumber \\
\omega_{0,1}^-&=e^{-\varepsilon_{A_-P}} \nonumber \\
\omega_{0,1}^+&=e^{-\varepsilon_{A_+P}} \nonumber \\
\omega_{b,1}^+&=e^{-\varepsilon_{A_-A_+}^{b,1}} \nonumber \\
\omega_{b,1}^-&=e^{-\varepsilon_{A_-A_-}^{b,1}} .
\label{eq:parameters}
\end{align}

Thus, the probability of RNA polymerase being bound to the
\textit{dnaA} promoter is

\begin{align}
\mathbb{P}(\text{RNAP bound})=\frac{1}{Z_{tot}} &\left(
Z(P-1,A_-,A_+)e^{-\varepsilon_{Pd}^{S}}+Z(P-1,A_--2,A_+)e^{-\varepsilon_{Pd}^{S}}k_{1,2}^-\omega_{0,1}^-\right.
\nonumber \\
&\left.+Z(P-1,A_-,A_+-2)e^{-\varepsilon_{Pd}^{S}}k_{1,2}^+\omega_{0,1}^+\right),
\end{align}

i.e.

\begin{equation}
\mathbb{P}(\text{RNAP bound})=\frac{1}{1+e^{\Delta
\varepsilon_{Pd}}\frac{N_{NS}}{PF_{reg}}},
\end{equation}

where 

\begin{equation}
F_{reg}^{-1}=\frac{1+\Delta
k_{b,c}\left(\frac{A_-}{N_{NS}}\right)^2+\Delta
k_{1,2}^-\left(\frac{A_-}{N_{NS}}\right)^2+\Delta
k_{1,2}^+\left(\frac{A_+}{N_{NS}}\right)^2+\Delta k_{b,c}\Delta
k_{1,2}^-\omega_{b,1}^-\left(\frac{A_-}{N_{NS}}\right)^4+\Delta
k_{b,c}\Delta k_{1,2}^+\frac{A_-^2A_+^2}{N_{NS}^4}}{1+\Delta
k_{1,2}^-\omega_{0,1}^-\left(\frac{A_-}{N_{NS}}\right)^2+\Delta
k_{1,2}^+\omega_{0,1}^+\left(\frac{A_+}{N_{NS}}\right)^2},
\end{equation}

where it has been assumed that $N_{NS}\gg P$, $A_-$ and $A_+$.
In this equation, $\Delta k_{i,j}^{(+/-)}=k_{i,j}^{(+/-)}\times
e^{2(\varepsilon^{NS}_{A_{+/-}d})}$.

Now, if the substitution $r\kappa=\frac{A_-}{N_{NS}}$ is used (see main text
for definition of $\kappa$), then when a transformation such as that in~\ref{sec:math_trans}, namely

\begin{equation}
r'=\lambda r - a,
\end{equation}

is attempted, it can be seen that the $(r\kappa)^2$ and $(r\kappa)^4$ terms will create
terms that are linear, and cubic, in $r'$.
Thus, it is not possible in general to make a scaling and translation
on $r$ in this way and keep the $F_{reg}$ in the same form.
This information shows that this model of the promoter, with fixed parameters,
cannot explain a constant threshold for the ratio $r$ at initiation, and
suggests that analyzing this promoter poses a different problem
altogether as it cannot be done by applying the techniques used in
this work for the more simple promoter.
Moreover, equation (\ref{eq:parameters}) introduces many new
parameters in the model, which generally cannot be directly estimated
from experiments, and therefore adds an element of uncertainty into
the model. For this reason it is believed that, while the promoter model
is satisfactory with respect to existing footprinting data~\cite{Speck1999}
, in absence of more precise knowledge, a simple
controlled model such as the one presented in the main text is to be
preferred.

\section{Considerations on other model ingredients}
 \subsection{DnaA-ATP recycling regions}
 \label{sec:DARS_considerations}
 \rev{As explained in the main text, the DnaA recycling sequences,
   known to convert the ADP-bound form of DnaA into its ATP-bound form
   were considered in a model variant. Here we discuss the mathematics
   of including a term in the equation to represent this reactivation.
   We show how it counters RIDA, with a rate that is proportional to
   the genome amount rather than genome replication rate, and analyze
   its effects under variations of the growth rate.}
 
 \rev{The number of the DnaA recycling sites is assumed to be
   proportional to the total length of the genome in the
   cell. Furthermore, regeneration is assumed to be replication fork
   independent, and it is assumed that the rate limiting parameter is
   the recycling rate per recycling site, rather than the amount of
   DNA-ADP in the cell. Thus the new, effective, RIDA term becomes:}
 
 \begin{equation}
 	\tilde{k_R} = k_R -  \rho \frac{\Lambda}{\mathcal{F}},
 \end{equation} 
 \rev{ where $\rho$ is the recycling rate per length of chromosome
   (i.e. $\rho$ is (the recycling rate per site)$\times$(the number of
   recycling sites per genome equivalent). Performing the same
   transformation as in equation \eqref{eq:transformation} gives }
 
 \begin{equation}
 	k_R' = \lambda k_R + a k_\Lambda
 \end{equation}
 
 and
 \begin{equation}
 	\rho'=\lambda \rho
 \end{equation}
 
 \rev{i.e. unless $\lambda=1$, the recycling rate per site must also
   vary with growth rate. This indicates that the effect of DARS
   recycling is not able by itself to impose a constant threshold with
   varying growth rate. }

\rev{Note that one now has}
 \begin{equation}
 	\tilde{k}_R' = \lambda \tilde{k}_R + a k_\Lambda \ ,
\end{equation}

\rev{ which is equivalent to equation \eqref{eq:kBdash} with
  $\tilde{k}_R$ in place of $k_R$. Thus this new, effective RIDA rate
  behaves similarly as the previous RIDA rate does in the main
  equation, scaling and translating as a function of cell doubling
  time.  }

\subsection{Specific binding sites for DnaA}
\label{sec:specific_binding}
\rev{This section considers a model variant including specific sites
  along the chromosome, at which DnaA-ATP can bind, providing a
  titrating effect reducing the reservoir of free DnaA-ATP that can be
  bound non-specifically on the chromosome. We show that the inclusion
  of a term of this form is equivalent to a reduction of the RIDA
  rate.}

\rev{Let $\beta$ denote the number of specific binding sites per
  genome equivalent. We further assume that once these specific
  binding sites are created, they are (nearly) always bound by a
  DnaA-ATP molecule. Thus}

 \begin{align}
   \frac{\partial A_- }{\partial t} &= Q - k_R\mathcal{F} - \beta
   \frac{\partial \Lambda}{\partial t} \nonumber \\ 
   &=Q - (k_R + \beta k_\Lambda) \mathcal{F},
 	\label{eq:titration}
\end{align}

\rev{where $A_-$ now represents the `free' DnaA-ATP. Thus, equation
  \eqref{eq:titration} is the same as the main equation for
  $\frac{\partial A_-}{\partial t}$, but with an effective increase of
  the RIDA rate due to titration, where the new, effective, RIDA rate
  is given by}

\begin{equation}
	\tilde{k}_R = k_R + \beta k_\Lambda.
\end{equation}

\rev{$\beta\approx 300$ and $k_\Lambda=\frac{1}{40}$min$^{-1}$, hence
  $\beta k_\Lambda \approx 7.5$min$^{-1}$. As shown in Additional
  Figure~\ref{fig:robustness1}, the model is robust to changes in the
  RIDA rate of this order of magnitude, and thus adding the role of
  specific binding sites does not affect the qualitative behavior of
  the model.}

\section{Experimental Methods}
\rev{
\label{sec:experimental_methods}
\emph{Escherichia coli} K-12 strains BW25113 carrying the
pKK-\emph{gfp} plasmid, where the reporter gene \emph{gfp} (green
fluorescent protein) was expressed under control of the \emph{dnaA}
promoter region (PdnaA), were grown at 37$^\circ$C in four different M9
minimal media to support different growth rates.  Fluorescence and
optical density were measured as a function of time with an automated
temperature-controlled plate reader Wallac Victor3. Gene expression
was calculated by taking the time derivative of the fluorescence
divided by the optical density, which provides a measure of the
promoter activity as previously described~\cite{Zaslaver2009}. }

\rev{
To obtain data for the promoter activity as a function of growth rate
during exponential phase, an exponential window was automatically
detected by taking as maximum value the inflection point of the OD
curve and as minimal value the first OD value that is two times higher
than the OD background value. In this window the data was then fitted
to an exponential function and this fit was used in order to extract
the values of generation time. Expression data for different growth
rates were corrected for plasmid abundance.  }

\newpage
\small

\newpage

\section{Additional Figures}

\begin{figure}[!ht]
  \includegraphics[width=\textwidth]{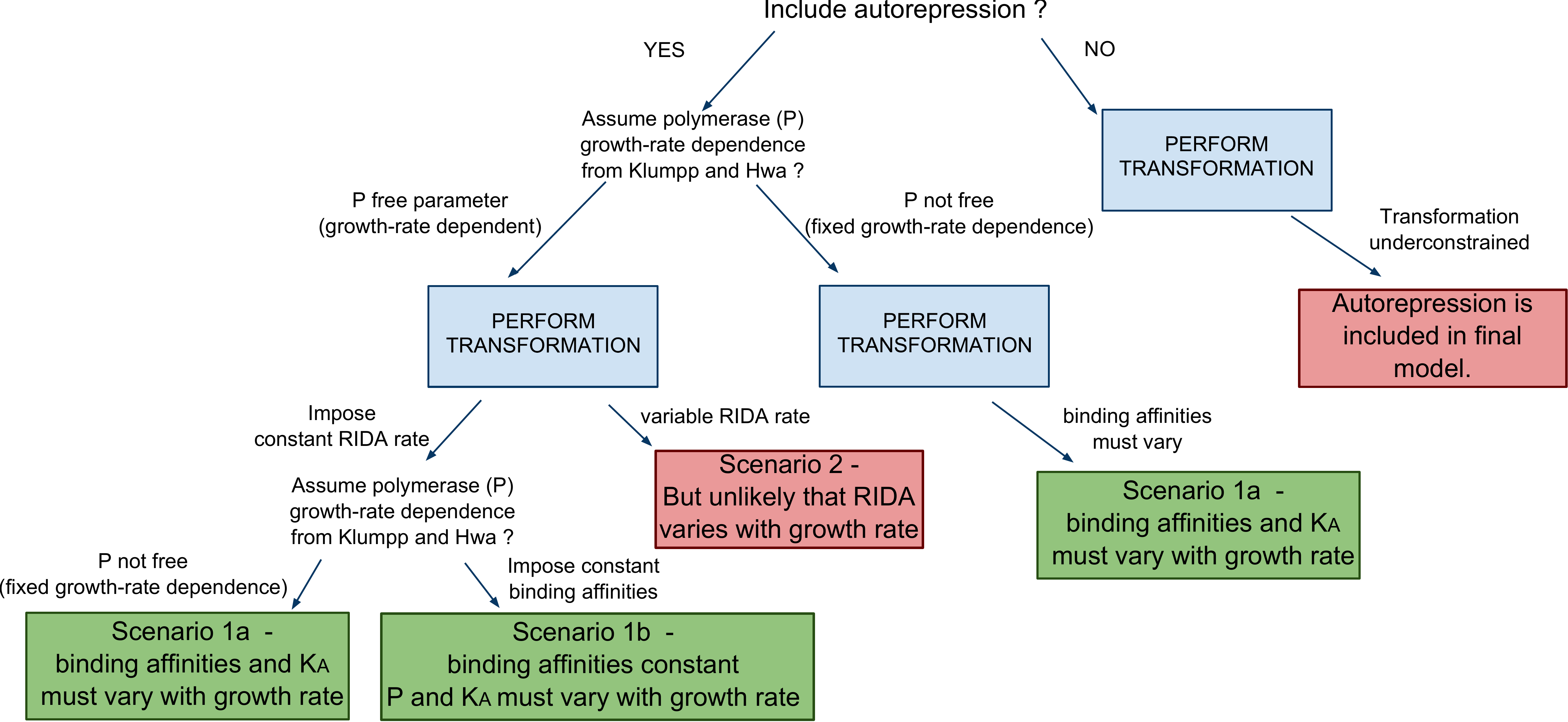}
	\caption{\textbf{Flow-chart of the procedure adopted for
            defining the scenarios of parameter variation with growth
            rate.}}
  \label{fig:SF0}
\end{figure}


\begin{figure}[!ht]
\begin{center}
	\includegraphics[width=0.7\textwidth]{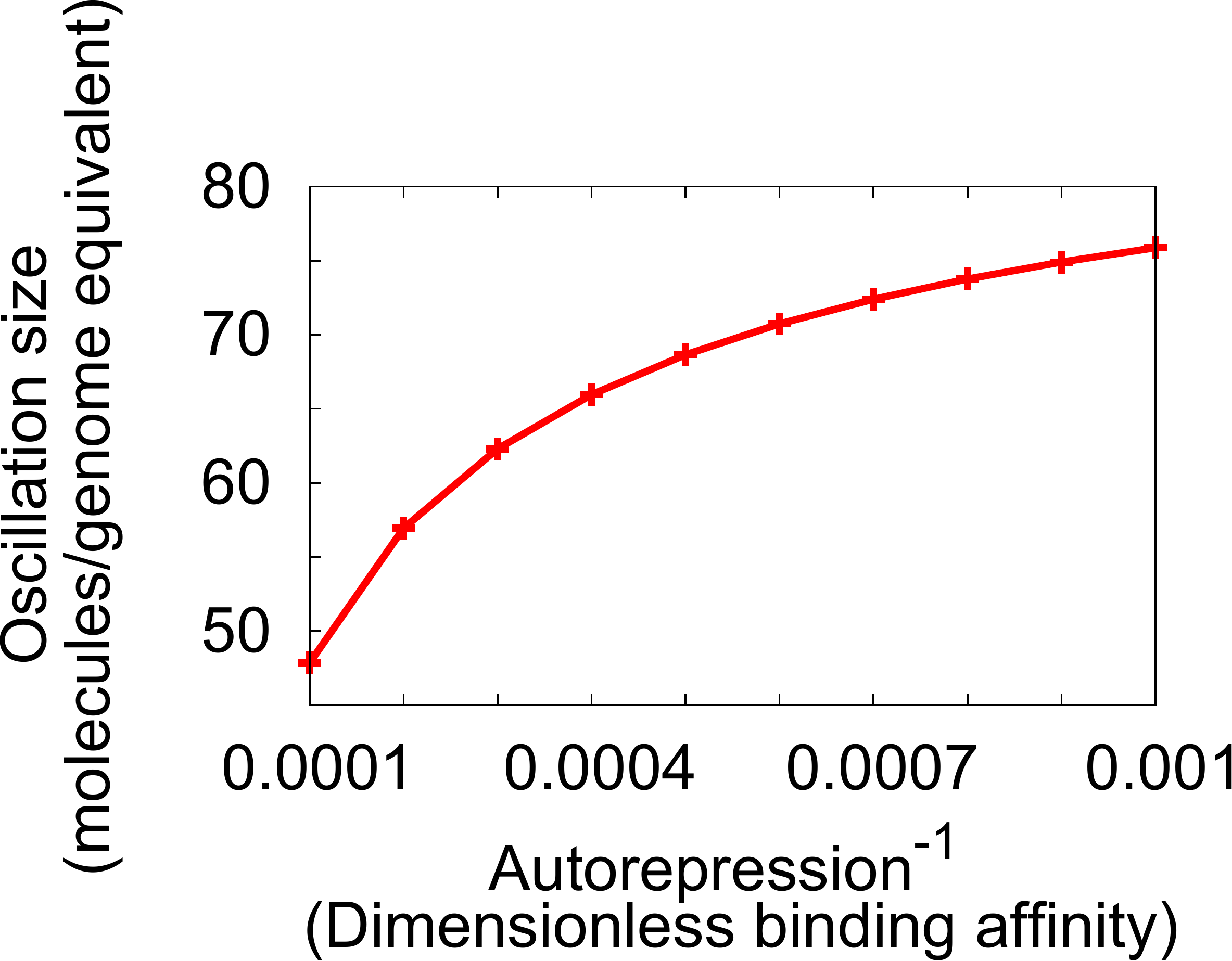}	
\end{center}
\caption{\textbf{Autorepression reduces the amplitude of the
    oscillations of the cell cycle.} If one varies the dimensionless
  binding affinity of DnaA-ATP to its self-repression sites, one can
  effectively vary the amount of autorepression. As the amount of
  autorepression is increased, the amplitude of the oscillations of
  the cell cycle decrease. The amplitude is defined as the difference
  between the maximum and minumum values of $r$ in a given cell
  cycle. This cell cycle length in this plot is $\tau=35$ mins. }
  	\label{fig:SF5}  	
\end{figure}

\begin{figure}[!ht]
\begin{center}
\includegraphics[width=0.7\textwidth]{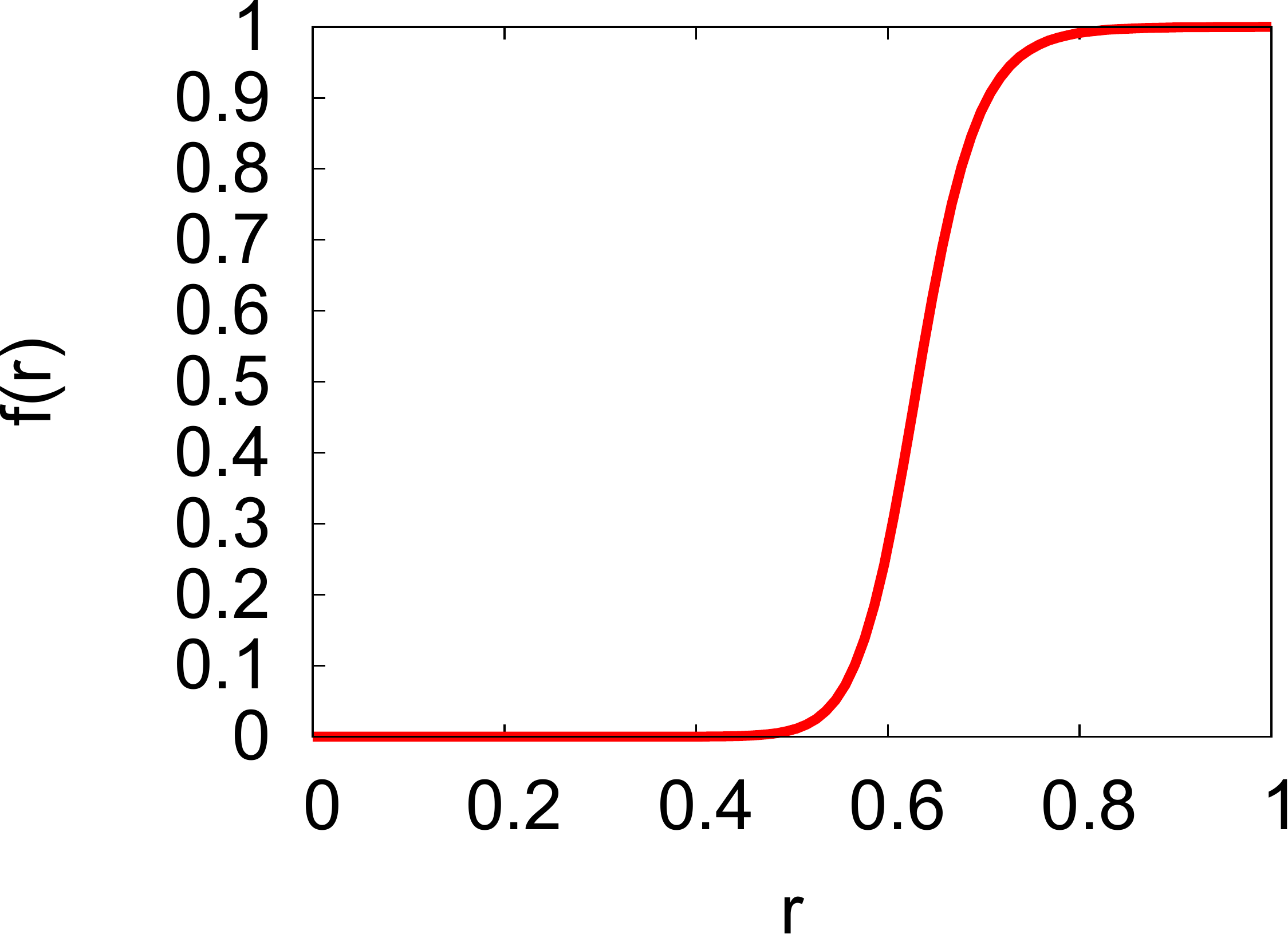}
\end{center}
\caption{\textbf{There is a critical value at which the probability of
    DnaA-ATP molecules binding to the origin quickly approaches
    unity.} In this plot, $f(r)=\frac{1}{\omega e^{20\Delta
      \varepsilon/k_B T}r^{-20}+1}$, which is the probability of
  having $20$ DnaA-ATP molecules bound to the origin, where we have
  taken $\omega e^{20\Delta \varepsilon/k_B T}=0.0001$ in this case.}
  \label{fig:SF10}
\end{figure}

\begin{figure}[!ht]
\includegraphics[width=\textwidth]{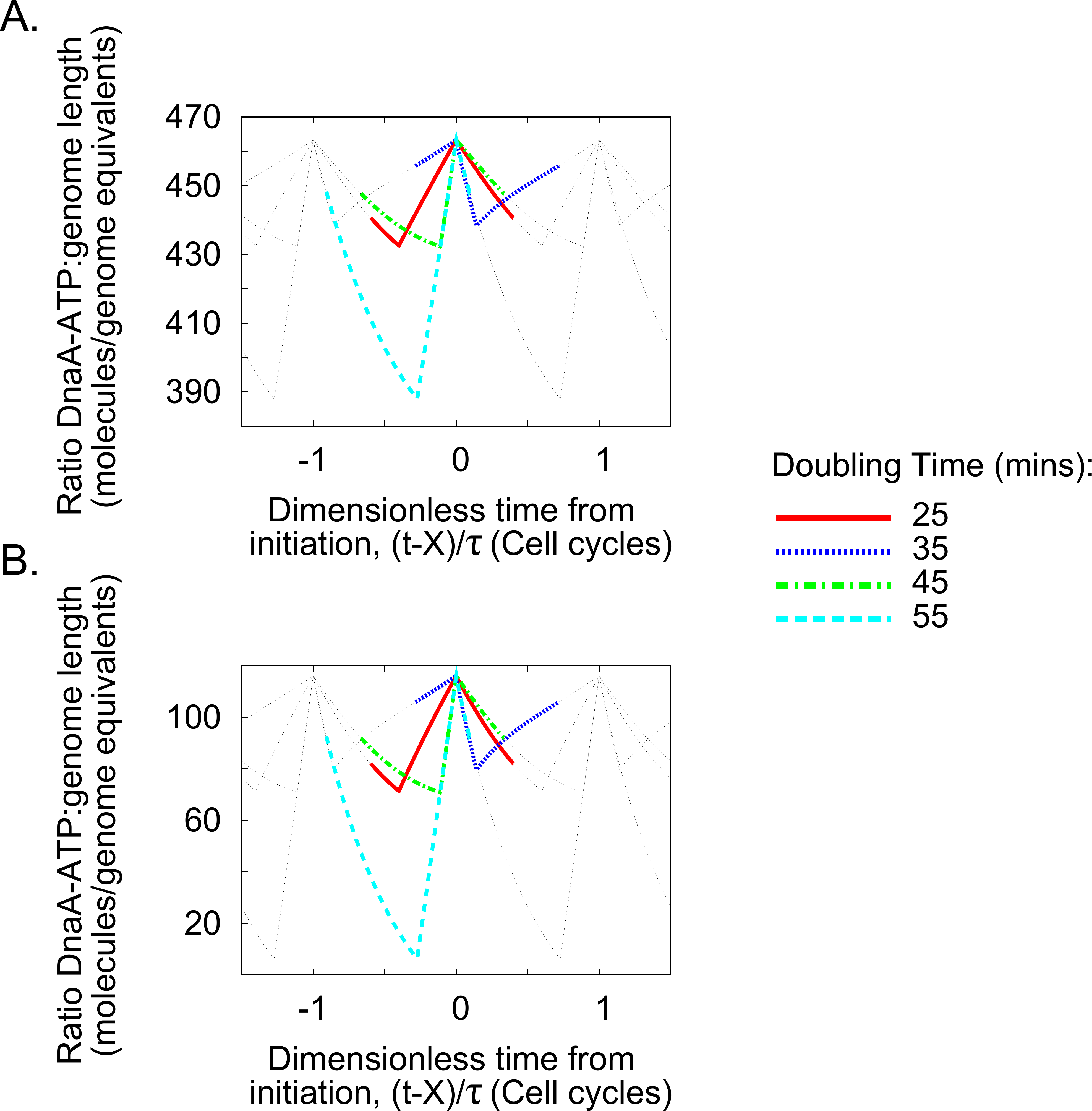}
\caption{\textbf{A constant threshold can still be achieved when the
    RIDA rate $k_R$ is varied significantly.} A: The oscillations of
  the cell cycle when $k_R=2$ molecules/minute. B: The oscillations of
  the cell cycle when $k_R=17$ molecules/minute. Despite RIDA rates
  being nearly 10 fold different, a constant threshold can still be
  achieved for each. }
  \label{fig:SF8}
\end{figure}

\begin{figure}[!ht]
\begin{center}
	\includegraphics[width=0.7\textwidth]{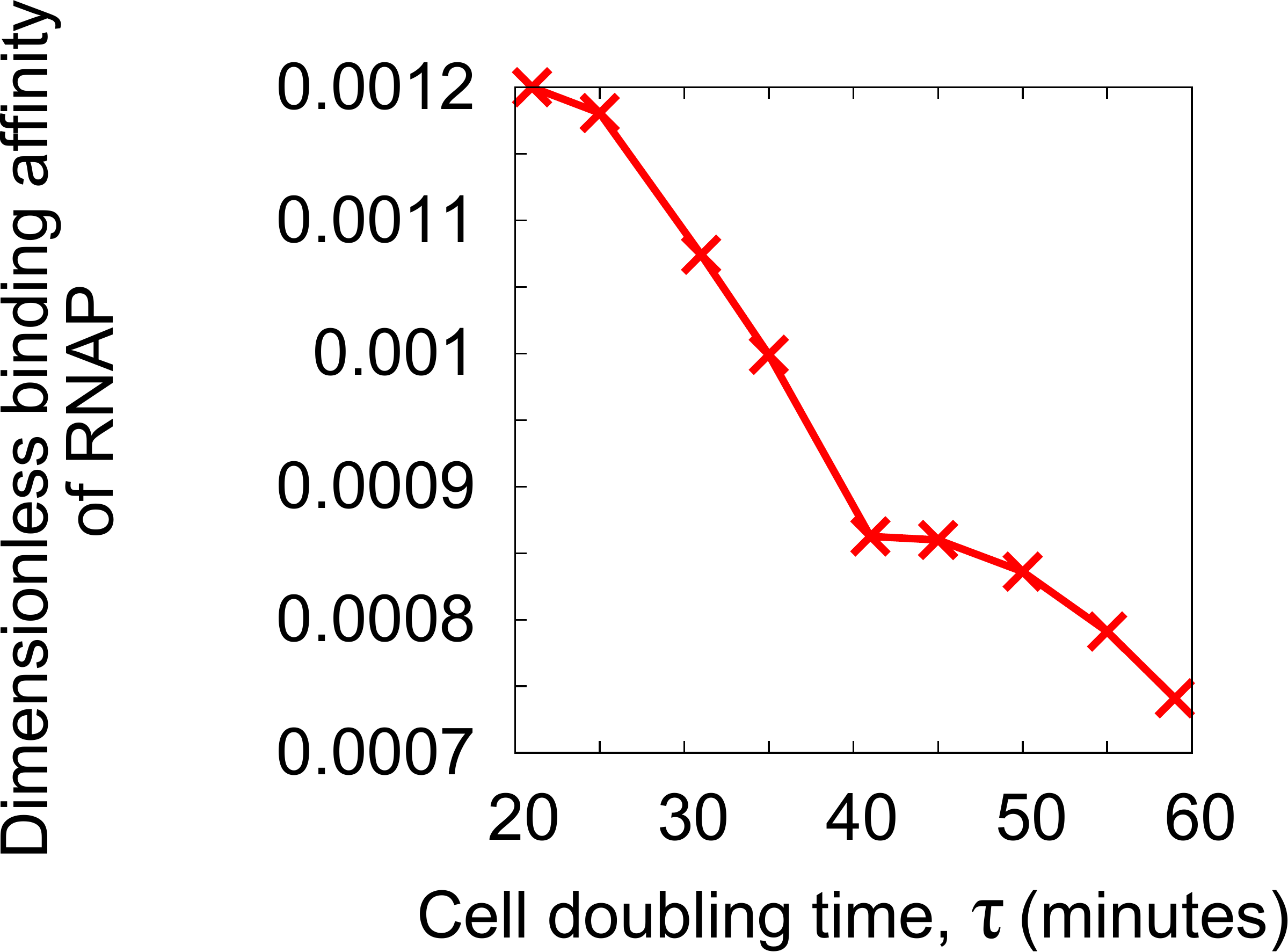}
\end{center}
\caption{\textbf{In scenario 1a, the binding affinity of RNAP to the
    DnaA promoter ($e^{\Delta \varepsilon_{pd}/k_BT}$) varies with
    cell doubling time ($\tau$).} In particular it decreases with cell
  doubling time, in a similar manner to the binding affinity of
  DnaA-ATP. This could be caused by supercoiling.}
  \label{fig:SF3}
\end{figure}

\begin{figure}[!ht]
	\includegraphics[width=\textwidth]{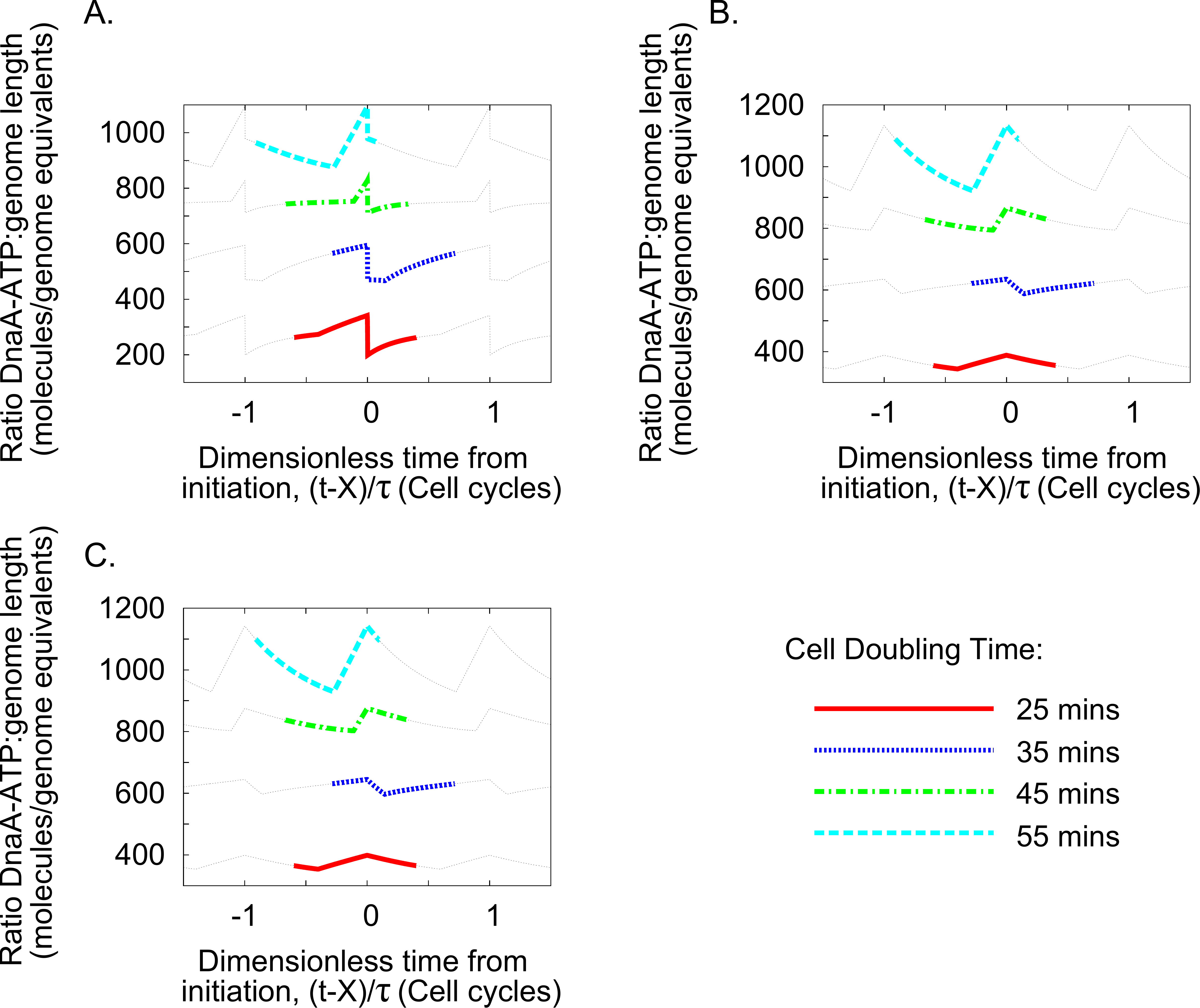}
	\caption{\textbf{Model variants were considered which all fail
            to explain a constant initiation threshold in $r$.} The
          models include: A: The \textit{datA} locus, binding up to
          300 DnaA-ATP molecules soon after initiation. B: A delay in
          the synthesis of DnaA-ATP, to reflect the time interval
          between the mRNA being transcribed and the DnaA protein
          being translated. C: A more complex form for the
          \textit{dnaA} promoter, containing two binding sites for
          DnaA which binds cooperatively.  None of these variants
          succeeds in achieving a constant initiation threshold in
          $r$, leading us to pursue a model in which some of the
          parameters of the model are able to vary with growth rate.}
  \label{fig:SF4}
\end{figure}

\begin{figure}[!ht]
\includegraphics[width=\textwidth]{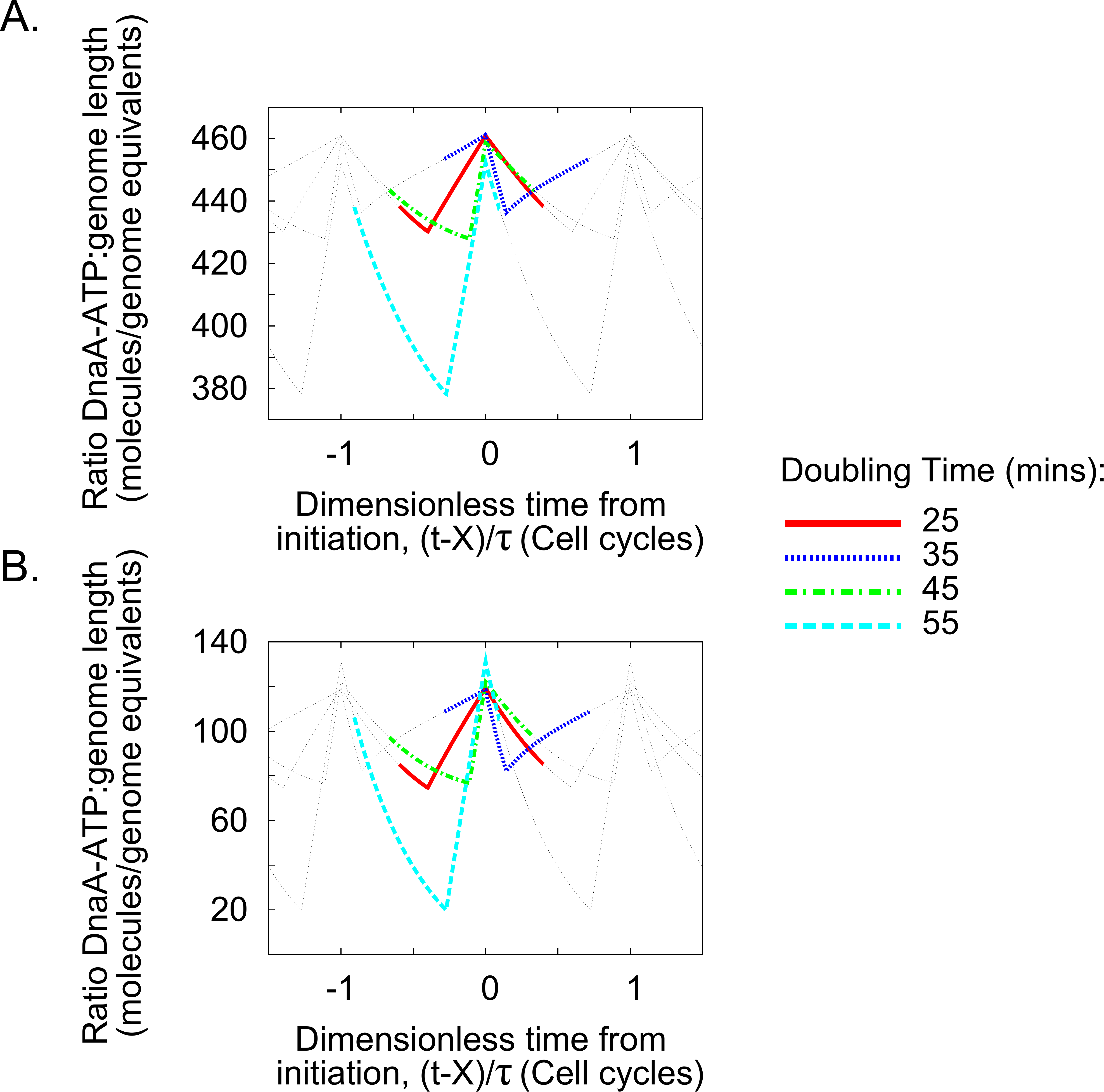}
\caption{\textbf{A mutant in which the RIDA rate is varied.}
  \rev{Attempting to reproduce the effect of over or under-expressing
    the Hda protein or sequestering DnaA from specific titration sites
    uniformly distributed along the genome, we modified the RIDA rate
    leaving the other parameters fixed to their values when $k_R=10$
    molecules/minute. A: The RIDA rate is reduced to $k_R=2$
    molecules/minute, resulting in a lower value of $r$ at $t=X$ at
    slower growth rates, and thus later initiation time. B: The RIDA
    rate is increased to $k_R=17$ molecules/minute, resulting in a
    higher value of $r$ at $t=X$ at slower growth rates, and thus
    earlier initiation time.} }
  \label{fig:SF9}
\end{figure}

\begin{figure}[!ht]
\vspace{-1cm}
\includegraphics[width=\textwidth]{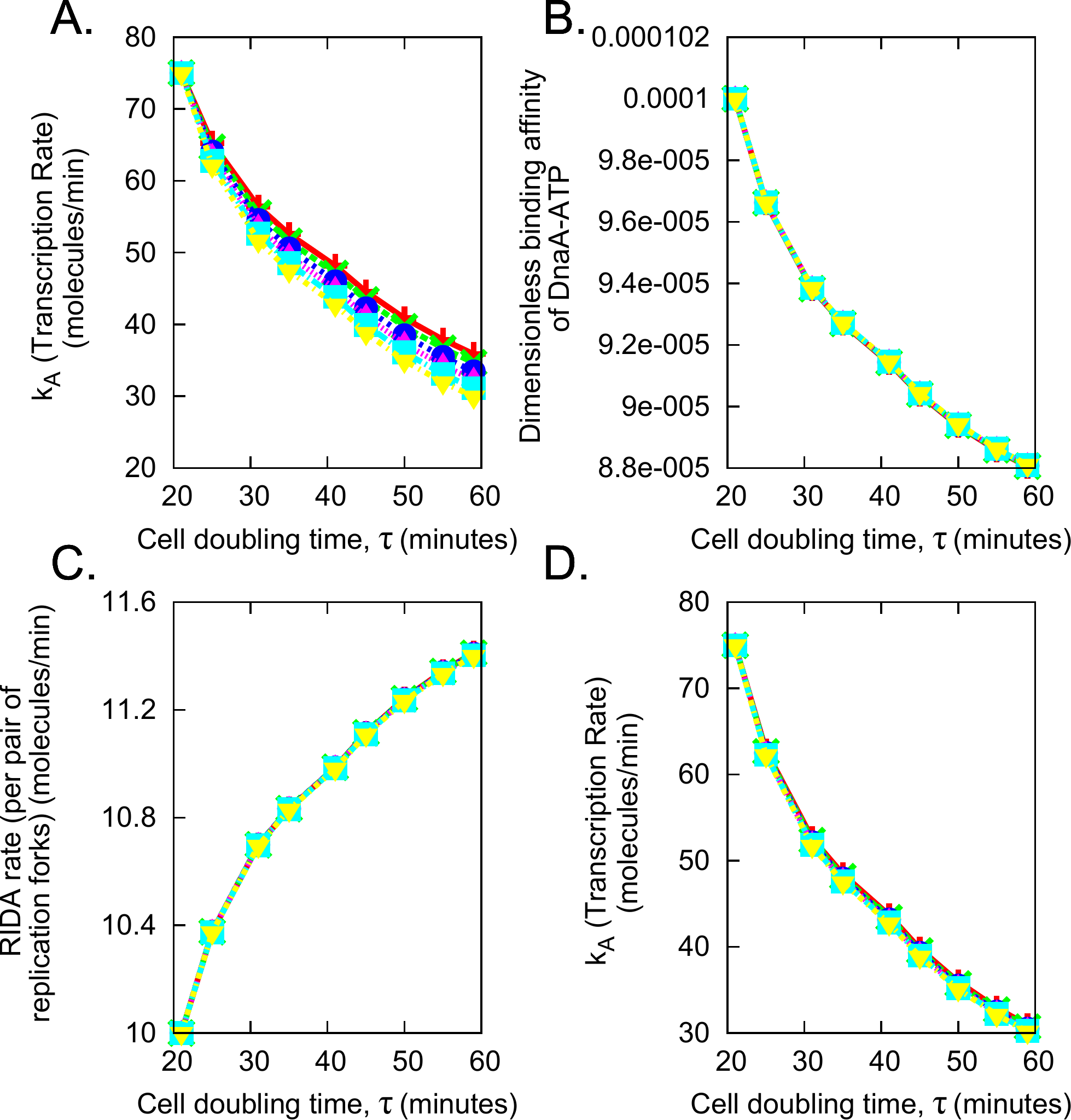}
\caption{\textbf{\rev{The model is robust to changes in the input
      parameters.}}  In order to show that the qualitative trends,
  observed from transforming the main equation, were independent of
  the choice of parameters, the values of the input parameters were
  varied and the trends replotted.  A: The trend of $k_A$ as the value
  of the RIDA rate, $k_R$, is varied, with $k_R \in \{5,
  7,9,11,13,15\}$, denoted by the symbols,
  \huge\textcolor{red}{\textbf{+}}\normalsize,
  \textcolor{green}{\textbf{\BigCross}},
  \textcolor{blue}{\FilledBigCircle},
  \textcolor{magenta}{\FilledBigTriangleUp},
  \textcolor{cyan}{\FilledBigSquare},
  \textcolor{yellow}{\FilledBigTriangleDown} respectively.  B: The
  trend of the dimensionless binding affinity of DnaA-ATP to its
  self-repression sites as $k_A$ is varied, with $k_A \in
  \{50,60,70,80,90,100\}$, denoted by the symbols,
  \huge\textcolor{red}{\textbf{+}}\normalsize,
  \textcolor{green}{\textbf{\BigCross}},
  \textcolor{blue}{\FilledBigCircle},
  \textcolor{magenta}{\FilledBigTriangleUp},
  \textcolor{cyan}{\FilledBigSquare},
  \textcolor{yellow}{\FilledBigTriangleDown} respectively.  C: The
  trend of the RIDA rate, $k_R$ as the dimensionless binding affinity
  of RNAP to the DnaA promoter is varied, with $e^{\Delta
    \epsilon_{pd}/k_BT} \in 10^{-4}\times\{5, 7, 9, 11, 13, 15\}$,
  denoted by the symbols, \huge\textcolor{red}{\textbf{+}}\normalsize,
  \textcolor{green}{\textbf{\BigCross}},
  \textcolor{blue}{\FilledBigCircle},
  \textcolor{magenta}{\FilledBigTriangleUp},
  \textcolor{cyan}{\FilledBigSquare},
  \textcolor{yellow}{\FilledBigTriangleDown} respectively.  D: The
  trend of $k_A$ as the dimensionless binding affinity of RNAP to the
  DnaA promoter is varied, with $e^{\Delta \epsilon_{pd}/k_BT} \in
  10^{-4}\times\{5, 7, 9, 11, 13, 15\}$, using the same symbols as in
  C. In all the plots, the qualitative trend is the same for all the
  parameter values.}
  \label{fig:robustness1}
\end{figure}

\begin{figure}[!ht]
\includegraphics[width=\textwidth]{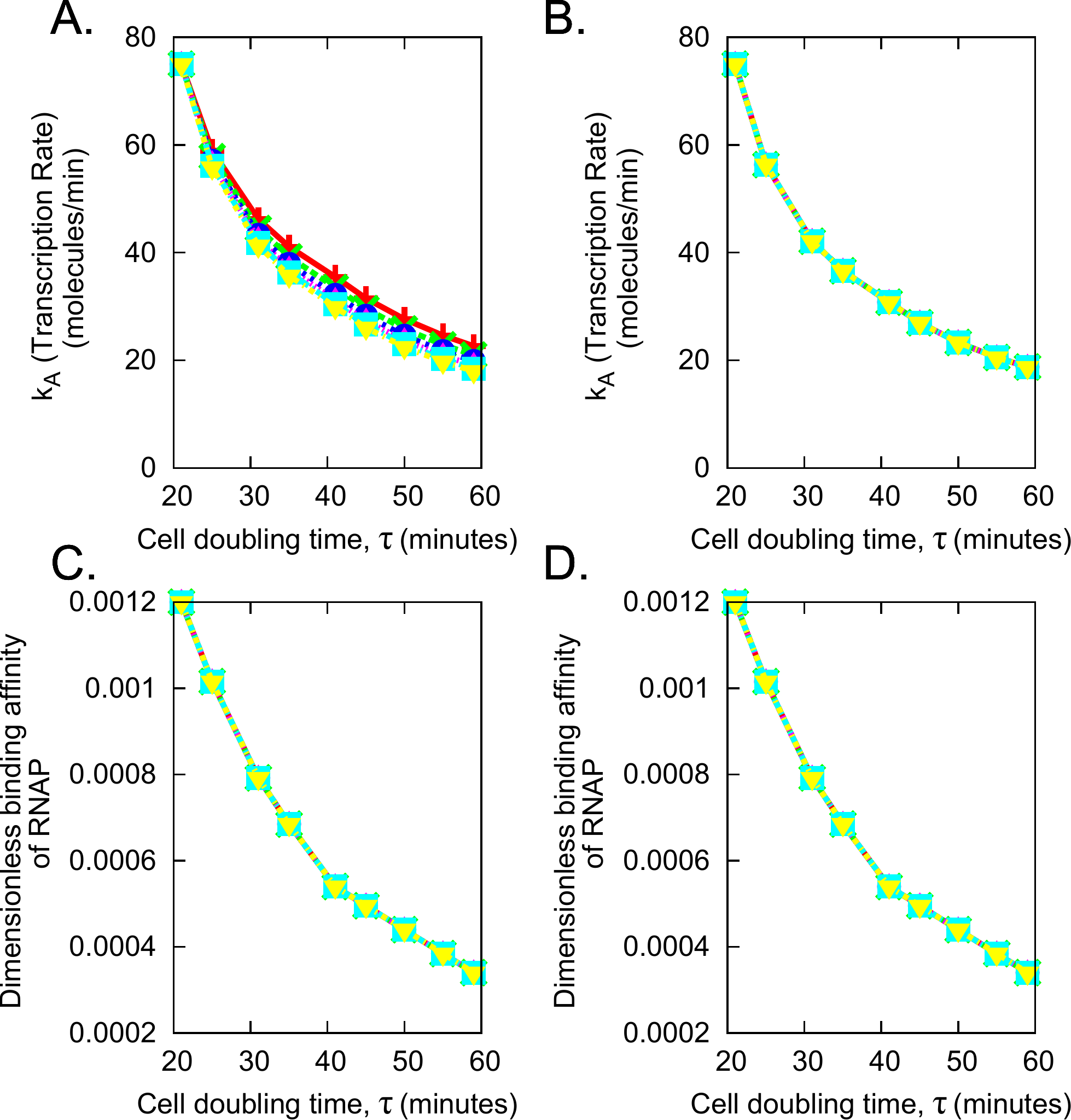}
\caption{\textbf{\rev{The model with no autorepression but with RIDA
      is robust to changes in the input parameters.}}  A: The trend of
  $k_A$ as the dimensionless binding affinity of RNAP to the DnaA
  promoter is varied. B: The trend of $k_A$ as the RIDA rate, $k_R$ is
  varied. C: The trend of the dimensionless binding affinity of RNAP
  to the DnaA promoter as $k_A$ is varied. D: The trend of the
  dimensionless binding affinity of RNAP to the DnaA promoter as the
  RIDA rate, $k_R$ is varied. In each case, the set different values
  taken by the varied parameter, and the corresponding symbols in the
  plots, are the same as those given in the caption of Additional
  Figure \ref{fig:robustness1}.}
  \label{fig:robustness2}
\end{figure}

\begin{figure}[!ht]
\includegraphics[width=\textwidth]{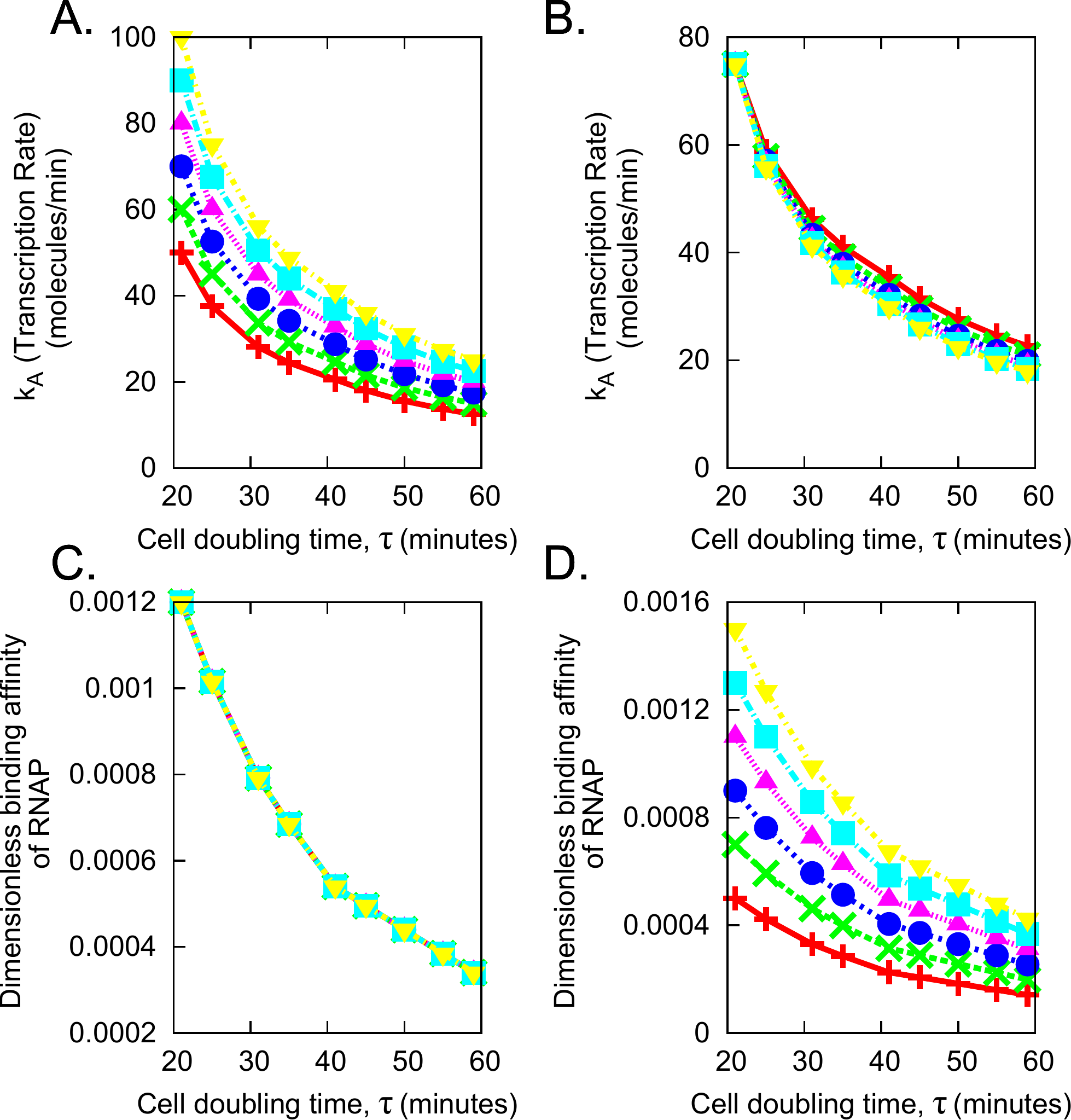}
\caption{\textbf{\rev{The model with no autorepression and no RIDA is
      robust to changes in the input parameters.}}  A: The trend of
  $k_A$ as the input value of $k_A$ is varied. B: The trend of $k_A$
  as the dimensionless binding affinity of RNAP to the DnaA promoter
  is varied. C: The trend of the dimensionless binding affinity of
  RNAP to the DnaA promoter as $k_A$ is varied. D: The trend of the
  dimensionless binding affinity of RNAP to the DnaA promoter is
  varied. In each case, the set different values taken by the varied
  parameter, and the corresponding symbols, are the same as those
  given in the caption of Figure
  \ref{fig:robustness1}. Note that the spread seen in A and D is due
  to the fact that these plots are relative to parameters whose
  initial values (at $\tau=21$mins) are themselves changed. The
  observation that the transformation makes the initial spread at
  $\tau=60$mins narrower for larger $\tau$ is further evidence of
  robustness. }
  \label{fig:robustness3}
\end{figure}

\begin{figure}[!ht]
\includegraphics[width=\textwidth]{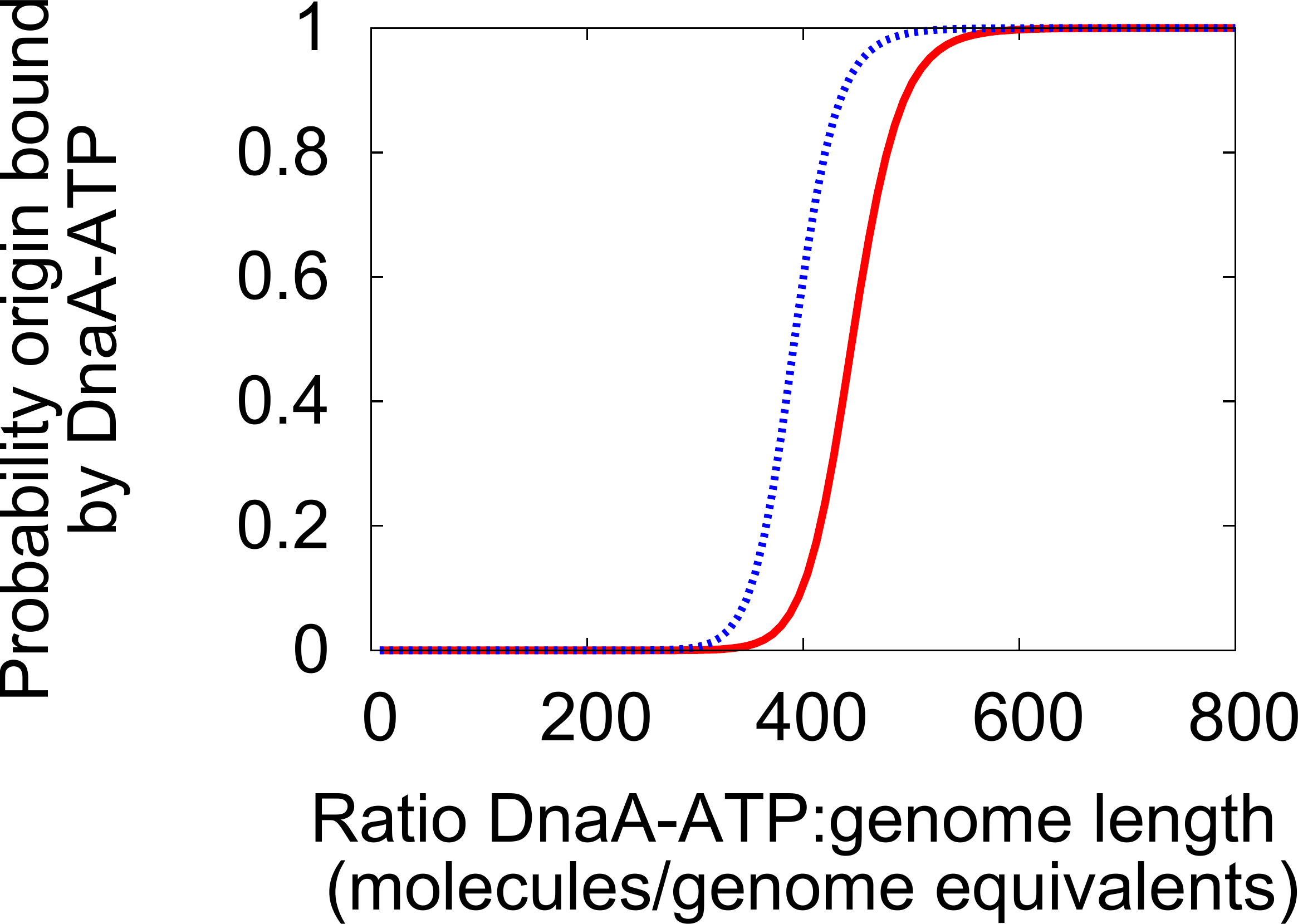}
\caption{\textbf{\rev{Varying the binding affinity of DnaA-ATP to the sites at the origin has only a minor effect on the initiation threshold.}} The probability of twenty DnaA-ATP molecules binding to the promoter, and hence starting initiation (as given in equation \eqref{eq:prob_origin_bound}) was plotted for different values of the binding affinity $e^{\Delta \varepsilon_{ao}/k_BT}$, to examine whether this had a large effect on the initiation threshold. The values chosen for the binding affinity were $e^{\Delta \varepsilon_{ao}/k_BT}= 0.0001$ (solid red curve), and $e^{\Delta \varepsilon_{ao}/k_BT}= 0.000088$ (dashed blue curve), which are the extreme values that the binding affinity of DnaA-ATP to its self repression sites attains in Scenario 1a (see Figure \ref{fig:fig4}). The initiation threshold is given by the inflection point of the curve. Thus, the difference between the initiation thresholds is $\approx 40$ (molecules/genome equivalents) $\approx 10 \%$ change. This is approximately constant, particularly when compared with the differences in $r(X)$ attained in the untransformed model (Figure \ref{fig:fig3}A), suggesting that this scenario might be robust.  }
  \label{fig:varying_threshold}
\end{figure}

\end{bmcformat}
\end{document}